\documentclass[aip,byrevtex,reprint,nobibnotes,floatfix,footinbib]{revtex4-1}

\bibpunct[, ]{[}{]}{,}{n}{,}{,}

\usepackage{setspace}
\usepackage{graphicx}
\usepackage{amssymb}
\usepackage{amsbsy}
\usepackage{amsmath}
\usepackage[varg]{txfonts}
\usepackage{psfrag}
\usepackage[colorinlistoftodos]{todonotes}
\usepackage{latexsym}
\usepackage{textcomp}
\usepackage{bm}
\usepackage[Euler]{upgreek}
\usepackage{dcolumn}
\usepackage[capitalize]{cleveref}

\usepackage{multirow}

\DeclareMathAlphabet{\mathsf}{OT1}{phv}{b}{n}

\newcommand{\crossVorg}{\ensuremath{%
         \setbox0=\hbox{$V$}
        V \kern-\wd0{\raise.3ex\hbox{$\relbar$}}}}

\newcommand{\crossVxx}[2]{%
	{\setbox0=\hbox{$#1#2V$}
         \setbox1=\hbox{$#1#2$}
         \setbox2=\hbox{$#1V$}
         \dimen1=\wd0
	 \advance\dimen1-\wd1
         \raise.2\ht0\hbox{$#1#2$}\kern-.4\wd0}}

\newcommand{\degree}{\ensuremath{^\circ}}
\newcommand{\degC}{\ensuremath{\degree\mathrm{C}}}

\usepackage{pifont}

\newcommand{\Rgth}{\ensuremath{R_{\mathrm{g}}^{\theta}}}

\newcommand{\Rg}{\ensuremath{R_{\mathrm{g}}}}

\newcommand{\Na}{\ensuremath{N_{\mathrm{A}}}}

\newcommand{\mk}{\ensuremath{m_{\mathrm{k}}}}

\newcommand{\kB}{\ensuremath{k_{\mathrm{B}}}}

\newcommand{\ag}{\ensuremath{\alpha_{\mathrm{g}}}}

\newcommand{\etas}{\ensuremath{\eta_{\mathrm{s}}}}
\newcommand{\etap}{\ensuremath{\eta_{\mathrm{p}}}}
\newcommand{\etapo}{\ensuremath{\eta_{\mathrm{p0}}}}

\newcommand{\leta}{\ensuremath{\lambda_{\eta}}}

\newcommand{\Ttheta}{\ensuremath{T_{\theta}}}

\newcommand{\ivisc}{\ensuremath{[\eta]}}

\newcommand{\Uer}{\ensuremath{U_{\eta \mathrm{R}}}}
\newcommand{\ccs}{\ensuremath{c / c^{*}}}

\newcommand{\nueff}{\ensuremath{\nu_{\mathrm{eff}}}}

\DeclareMathAlphabet\mathbfcal{OMS}{cmsy}{b}{n}
\DeclareMathAlphabet{\mathbfsf}{\encodingdefault}{\sfdefault}{bx}{sl}

\makeatletter 
\gdef\@ptsize{0}
\let\@currsize\normalsize 
\makeatother 
\usepackage{setspace} 

\begin{document}

\title{Shear thinning in dilute and semidilute solutions of polystyrene and DNA }


\author{Sharadwata Pan}
\affiliation{IITB-Monash Research Academy, Indian Institute of Technology Bombay, Powai, Mumbai - 400076, India}
\affiliation{Department of Chemical Engineering, Indian Institute of Technology Bombay, Powai, Mumbai - 400076, India}
\affiliation{Department of Chemical Engineering, Monash University, Melbourne, VIC 3800, Australia}

\author{Duc At Nguyen}
\affiliation{Department of Chemical Engineering, Monash University,
Melbourne, VIC 3800, Australia}

\author{B. D\"{u}nweg}
\affiliation{Department of Chemical Engineering, Monash University, Melbourne, VIC 3800, Australia}
\affiliation{Max Planck Institute
  for Polymer Research, Ackermannweg 10, 55128 Mainz, Germany}
\affiliation{Condensed Matter Physics, TU Darmstadt, Hochschulstra{\ss}e 12, 
  64289 Darmstadt, Germany}

\author{P. Sunthar}
\affiliation{Department of Chemical Engineering, Indian Institute of Technology Bombay, Powai, Mumbai - 400076, India}
\affiliation{IITB-Monash Research Academy, Indian Institute of Technology Bombay, Powai, Mumbai - 400076, India}

\author{T. Sridhar}
\affiliation{Department of Chemical Engineering, Monash University, Melbourne, VIC 3800, Australia}
\affiliation{IITB-Monash Research Academy, Indian Institute of Technology Bombay, Powai, Mumbai - 400076, India}

\author{J. Ravi Prakash}
\email[Corresponding author: ]{ravi.jagadeeshan@monash.edu}
\affiliation{Department of Chemical Engineering, Monash University, Melbourne, VIC 3800, Australia}
\affiliation{IITB-Monash Research Academy, Indian Institute of Technology Bombay, Powai, Mumbai - 400076, India}

\begin{abstract}
The viscosity of dilute and semidilute unentangled DNA solutions, in steady simple shear flow, has been measured across a range of temperatures and concentrations. For polystyrene solutions, measurements of viscosity have been carried out in the semidilute unentangled regime, while results of prior experimental measurements in the dilute regime have been used for the purpose of data analysis, and for comparison with the behaviour of DNA solutions. Interpretation of the shear rate dependence of viscosity in terms of suitably defined non-dimensional variables, is shown to lead to master plots, independent of temperature and concentration, in each of the two concentration regimes. In the case of semidilute unentangled solutions, defining the Weissenberg number in terms of a concentration dependent large scale relaxation time is found not to lead to data collapse across different concentrations. On the other hand, the use of an alternative relaxation time, with the concentration dependence of a single correlation blob, suggests the existence of universal shear thinning behaviour at large shear rates.

\end{abstract}


\maketitle

\section{\label{sec:intro} Introduction}

Two distinct regimes of behaviour are always observed in steady simple shear flows of polymer solutions  when the polymer contribution to viscosity \etap\ is plotted against the shear rate $\dot \gamma$,  regardless of the domain of concentration. At low shear rates, the solution is essentially Newtonian, with a constant viscosity independent of shear rate, typically denoted as the zero shear rate viscosity $\eta_{\text{p}0}$. Beyond a threshold value of shear rate, of magnitude roughly $\dot \gamma \approx \lambda^{-1}$, where $\lambda$ is a large scale relaxation time, the viscosity decreases with increasing shear rate, exhibiting a power law dependence on $\dot \gamma$ at sufficiently large values of shear rate~\cite{birdetal1}. In the case of dilute polymer solutions, some experimental and computational studies suggest the existence of a third regime at even higher shear rates, where the viscosity levels off and displays another Newtonian regime independent of shear rate~\cite{birdetal1,Noda:1968gt,Liu19895826,Doyle1997251}. Though the existence of the first two regimes has been observed in all experimental studies, and predicted in many analytical and computational studies, there are significant differences in the reported exponent for the power law decay of viscosity. The source of this discrepancy is probably the result of the experimental studies being conducted under widely varying  conditions of temperature ($T$), polymer molecular weight ($M$),  concentration ($c$), and ranges of shear rate~\cite{Kotaka:1966cl,Noda:1968gt,Suzuki:1969ec,Mochimaru:1983do,Hua2006787,Hur2001421,Teixeira:2005p1058,Schroeder20051967,Heo20088903}, and the theoretical studies being carried out with different polymer models and underlying assumptions~\cite{Stewart:1972gt,Fan:1985jk,Bird1987,Liu19895826,zylka:jcp-91,Doyle1997251,Petera19997614,Aust19995660,prakash:jor-02,Winkler2010}. The goal of the present work is to examine if it is possible to collapse the viscosity versus shear rate data, obtained experimentally for dilute and semidilute unentangled solutions  of polystyrene and DNA across a wide range of conditions, onto universal master plots by appropriately choosing  variables that represent the state of the solution when it is undergoing simple shear flow. It is anticipated that such a representation of the data would lead to a more unambiguous understanding of the nature of shear thinning itself, and of the power law  exponent. 

At equilibrium, for dilute and semidilute solutions of polymers that are neutral, flexible and linear, it has been established that just two variables, the solvent quality $z$ and the scaled concentration $c/c^*$, are sufficient to collapse the static and dynamic properties of a wide variety of different polymer-solvent systems, onto master plots~\cite{GrosbergKhokhlov94,Schafer99,RubCol03,Colby:2010uv,Jain2012a,Pan2014339}. The solvent quality $z$, which is the scaling variable that describes the temperature crossover behaviour from $\theta$ solvents to very good solvents,  combines the dependence on $T$ and $M$, while $c/c^*$ (where $c^*$ is the overlap concentration which determines the onset of the semidilute regime), combines the dependence on $c$ and $M$. In the presence of flow, a number of experimental and theoretical studies suggest that plotting $\etap/\eta_{\text{p}0}$ versus $\lambda \dot \gamma$ for different systems leads to a collapse of data. However, these studies are invariably limited to a particular concentration regime, and in most  cases do not attempt to examine equilibrium and nonequilibrium behaviour within a common conceptual framework. In this work, we examine if the representation of viscosity versus shear rate data in terms of $z$, $c/c^*$ and $\lambda \dot \gamma$ (with particular attention paid to the appropriate choice of $\lambda$), is sufficient to reveal the underlying universal behaviour in dilute and semidilute unentangled solutions of polystyrene and DNA. 

The literature on experimental, analytical and computational studies of the shear thinning of polymer solutions is very extensive, and is spread over several decades in the past. For the purpose of placing the present work in context, we discuss a few papers that highlight the essential aspects of the experimental observations and theoretical predictions that have been made over the years, starting with dilute solutions before considering semidilute unentangled polymer solutions. 

Arguably the most careful experimental measurements of the shear thinning of dilute polymer solutions were carried out nearly 50 years ago with the help of capillary viscometers. Representative examples of this work are the  papers by~\citet{Kotaka:1966cl,Noda:1968gt} and~\citet{Suzuki:1969ec}, in which the intrinsic viscosity $[\eta]$ was measured as a function of the nondimensional shear rate $\lambda_Z \dot \gamma$, where, 
\begin{equation}
\label{eq:lambdaZ}
\lambda_Z = \frac{M [\eta]_0 \eta_s}{N_A k_B T}
\end{equation} 
is a large scale relaxation time based on the zero shear rate value of the intrinsic viscosity $[\eta]_0$. Here, $\eta_s$ is the solvent viscosity, $N_A$ is the Avogadro number and $k_B$ is Boltzmann's constant. \citet{Kotaka:1966cl} considered the behaviour of polystyrene solutions with polymers of several different molecular weights in four different solvents covering the range from $\theta$ to good solvents. Their experiments were carried out up to a maximum value of nondimensional shear rate $\lambda_Z \dot \gamma \approx 5$. They observed a collapse of data when $[\eta]/[\eta]_0$ was plotted against $\lambda_Z \dot \gamma$, for all the different molecular weights, temperatures  and solvents, regardless of the quality of the solvent.  They did not comment on the shear thinning exponent, nor did they observe the high shear rate constant viscosity regime. \citet{Noda:1968gt} measured the ratio $[\eta]/[\eta]_0$ for several different molecular weight samples of poly($\alpha$-methyl-styrene), and one polystyrene sample, in both $\theta$ and good solvents, for nondimensional shear rates up to a maximum value of $\lambda_Z \dot \gamma \approx 40$. In contrast to the observation by \citet{Kotaka:1966cl}, they observe a large difference in the behaviour  of the polystyrene sample in good and $\theta$ solvents, for values of $\lambda_Z \dot \gamma > \mathcal{O}(1)$. The thinning observed in good solvents is significantly greater than in $\theta$ solvents, with the viscosity in the $\theta$ solvent levelling off to a constant value for $\lambda_Z \dot \gamma > 5$. With regard to the dependence on molecular weight, they find that their data for the 13.6 $\times$ 10$^{6}$ g/mol (13.6 M) polystyrene sample in decalin (which is a $\theta$ solvent) exhibits less shear thinning than the 7.14 $\times$ 10$^{6}$ g/mol (7.14 M) polystyrene sample of~\citet{Kotaka:1966cl} in the same solvent, with a delayed onset of shear thinning. They also observe a clear influence of molecular weight in their observations of the shear thinning of the poly($\alpha$-methyl-styrene) samples, with an increase in the threshold value of $\lambda_Z \dot \gamma$ for shear thinning with increasing molecular weight. They report that their data does not permit them to extract a shear thinning exponent, and interestingly, they do not observe a collapse in any of their data sets. On the other hand, \citet{Suzuki:1969ec} observe data collapse for several of the polystyrene solutions considered by them when they plot $[\eta]/[\eta]_0$ against $\lambda_Z \dot \gamma$. For instance, they find that data for the 6.2 $\times$ 10$^{6}$ g/mol (6.2 M) polystyrene sample in three different solvents, all at their respective $\theta$ temperatures, collapse onto a single curve, for nondimensional shear rates up to a maximum value of $\lambda_Z \dot \gamma \approx \mathcal{O}(2)$. Similarly to~\citet{Noda:1968gt}, they find that increasing the solvent quality (while keeping the molecular weight fixed), increases the degree of shear thinning. However, in direct contrast to the observation of~\citet{Noda:1968gt}, they find that an increase in the polymer molecular weight (by roughly a factor of 2) leaves the onset of shear thinning relatively unchanged, but increases the degree of shear thinning. It should be noted that these authors also do not comment on the shear thinning exponent, and do not observe the levelling off in the viscosity at high shear rates, for the range of shear rates examined by them. 

By plotting $\eta_\text{p}/\eta_{\text{p}0}$ versus $\lambda^\dagger \dot \gamma$, \citet{Mochimaru:1983do} has shown that data reported previously in the literature for an extraordinarily broad range of polymer-solvent systems, across a wide range of molecular weights and concentrations (and a relatively narrower range of temperatures), can be collapsed onto a master plot if the large scale relaxation time $\lambda^\dagger$ is chosen appropriately. In particular, \citet{Mochimaru:1983do} has shown that $\lambda^\dagger$ can be chosen for each of these systems (regardless of the regime of concentration) such that experimental shear thinning data can be shifted onto a single curve that corresponds to the theoretical prediction of shear thinning by the FENE-P dumbbell model for dilute polymer solutions~\cite{Bird1987} (with the value of the finite-extensibilty parameter $b =50$). The experimental data typically have an upper limit to the nondimensional shear rate in the region $10^2$ to $10^3$.  Since the limiting value of the power law exponent for the FENE-P dumbbell model is $-(2/3)$, presumably all the experimental data considered in Ref.~\cite{Mochimaru:1983do} that lie on the theoretical curve conform to this exponent. It is worth mentioning here that~\citet{Bird1987}, in their monograph on the kinetic theory of polymeric liquids, have compiled an extensive set of experimental data published previously, in order to compare experimental measurements with theoretical predictions of shear thinning by the FENE and FENE-P dumbbell models for dilute polymer solutions.

The experiments of~\citet{Hua2006787} represent a relatively recent examination of the shear thinning of dilute polystyrene solutions in which  (rather than the intrinsic viscosity) the shear rate dependence of $\eta_p$ at finite concentrations below the overlap concentration $c^*$, has been considered. Measurements were carried out in a Couette viscometer for shear rates up to roughly $10^4$ s$^{-1}$ on polystyrene samples with four different molecular weights in dioctyl phthalate solutions at different temperatures that span $\theta$ and good solvents. \citet{Hua2006787} find that data collapse, independent of temperature, can be achieved if the ratio $\eta_\text{p}/\eta_{\text{p}0}$ is plotted versus $\lambda^* \dot \gamma$, where $\lambda^* = 1/\dot \gamma^*$, with $\dot \gamma^*$ being the threshold shear rate for the onset of shear thinning. When plotted in this manner, they find that the smaller molecular weight solutions exhibit a higher degree of shear thinning. The highest molecular weight solution is found to exhibit a power law exponent equal to $-(2/3)$. As a result, the data appears to suggest that the limiting value of $(2/3)$ for the magnitude of the shear thinning exponent, is approached from above with increasing molecular weight. The highest molecular weight solution also clearly exhibits a second Newtonian plateau value for $\eta_\text{p}$, for shear rates $\dot \gamma \gtrsim 10^4$. 

The use of single molecule experiments to observe fluorescently stained DNA in order to understand the relationship between microscopic dynamics and macroscopic rheology has been pioneered by the groups of Shaqfeh and Chu~\cite{Hur2001421,Teixeira:2005p1058,Schroeder20051967,Shaqfeh20051,Schroeder18}. In particular, the observation of molecular behaviour in the flow-gradient plane has led to significant new insights into the stretching and tumbling dynamics of individual molecules in simple shear flow. By making use of the relationship between the intensity of the measured image  and polymer density, \citet{Teixeira:2005p1058} and~\citet{Schroeder20051967} infer the accessible elements of the radius of gyration tensor. Since the Giesekus expression relates moments of the polymer conformation to the stress tensor, this enables them to estimate the dependence of viscosity on shear rate. Strictly speaking, the Kramers expression should be used to determine the stress tensor, since the Giesekus expression is not valid in the presence of hydrodynamic interactions~\cite{Bird1987}. However, \citet{Schroeder20051967} show through Brownian dynamics simulations which incorporate hydrodynamic interactions (discussed in more detail below), that the difference between predictions of viscosity by the Giesekus and Kramers expressions are not significant. By defining the Weissenberg number by $W\!i = \lambda_1 \dot \gamma$, where $\lambda_1$ is the longest relaxation time (obtained by an appropriate analysis of the relaxation of initially stretched molecules), both~\citet{Teixeira:2005p1058} and \citet{Schroeder20051967} observe that for Weissenberg numbers in the range $70 < W\!i < 1000$, the viscosity decays according to the power law, $\eta_\text{p} \sim W\!i^{-\alpha}$, with $\alpha=0.54$ reported by the former, and $\alpha=0.53$ by the latter.

Analytical attempts to predict the dependence of viscosity on shear rate for dilute polymer solutions date back all the way to the early development of kinetic theories for polymer solutions. Here we summarise some of these efforts, paying attention to the relationship between the underlying polymer model and the resultant prediction of shear thinning. 

Using a rigid dumbbell model for the polymer molecule with hydrodynamic interactions included, \citet{Stewart:1972gt} used the Galerkin method to predict the ratio $\eta_\text{p}/\eta_{\text{p}0}$ as a function of shear rate. They found that beyond the initial Newtonian plateau, the viscosity decreases with increasing shear rate, and approaches the asymptotic power law exponent $-(1/3)$ as $\dot \gamma \to \infty$. On the other hand, \citet{Fan:1985jk} used the Galerkin method to show that for a FENE dumbbell model, at moderate shear rates, the curve of $\log \left([\eta]/[\eta]_0\right)$ versus  $\log \left( \lambda_Z \dot \gamma \right)$ has a slope $-0.607$. In the special case of the FENE-P dumbbell model, it can be shown that the asymptotic power law exponent is equal to $-(2/3)$ as $\dot \gamma \to \infty$~\cite{Bird1987}. Thus the replacement of the rod with a spring as the connector between the beads leads to a significant change in the predicted value of the shear thinning exponent. None of these dumbbell models predict the second Newtonian plateau at high shear rates. 

An asymptotic power law exponent of $-(2/3)$ in the limit of large Weissenberg numbers was also observed by~\citet{Winkler2010}, who developed a model for dilute solutions of semiflexible polymers with the polymer molecules represented by a continuous differentiable space curve (with a fixed contour length), and hydrodynamic interactions incorporated in a pre-averaged manner. Curiously, \citet{Winkler2010} found that chain stiffness had very little influence on the dependence of viscosity on shear rate. 

A qualitatively different prediction of the dependence of viscosity on shear rate  is obtained when hydrodynamic interactions are treated with more sophisticated approximations than the simple pre-averaging approximation introduced originally by Zimm~\cite{Zimm1956269} in the context of a coarse-grained bead-spring chain model for dilute polymer solutions. A detailed discussion of a hierarchy of such models is given in the review article by~\citet{prakash:adv-99}. The most advanced approximations account for fluctuations in hydrodynamic interactions~\cite{Ottinger:1989up,prakash:jnnfm-97}, with a recent contribution accounting for fluctuations in both hydrodynamic interactions and in the FENE spring force law~\cite{prabhakar:jor-06}. For sufficiently long chains, the shape of the viscosity versus shear rate curve predicted by these models is rather complex. When a Hookean spring force law is used, after the initial Newtonian behaviour at low shear rates, there is a relatively weak shear thinning regime over a limited range of shear rates (with no discernible power law), followed by shear thickening for a range of shear rates (whose extent depends on chain length), until a constant plateau value is reached at high shear rates. The occurrence of shear thinning followed by shear thickening predicted by these closure approximations for hydrodynamic interactions has been confirmed subsequently by exact Brownian dynamics simulations~\cite{zylka:jcp-91,prabhakar:jor-06}. The use of a FENE spring force law leads to identical behaviour, except that subsequent to the onset of shear thickening, instead of reaching a high shear rate plateau value, a second shear thinning regime is observed which continues indefinitely for increasing shear rates, and approaches power law behaviour with an exponent equal to $-(2/3)$~\cite{wedgewood:jnnfm-88}. The initial shear thinning has been attributed to the presence of hydrodynamic interactions, while the second shear thinning regime has been associated with the finite extensibility of the chain. 

A renormalisation group calculation by~\citet{ott90}, based on a bead-spring chain model for the polymer has shown that just the presence of excluded volume interactions alone, without either finite extensibility or hydrodynamic interactions, leads to the prediction of shear thinning in dilute polymer solutions, with a power law exponent equal to $-0.25$ in the limit of high shear rates. A similar exponent has been predicted by~\citet{prakash:jor-02}, who accounted for fluctuating excluded volume interactions with the help of a Gaussian approximation. These approximate predictions of the shear thinning exponent in the presence of excluded volume interactions have been validated by Brownian dynamics simulations for bead-spring chains in the long chain limit~\cite{kumar:jcp-04}. 

Computer simulations of the behaviour of dilute polymer solutions in shear flow, with a view to predicting viscometric functions, have been carried out with coarse-grained bead-rod and bead-spring chain models, both with and without the inclusion of hydrodynamic and excluded volume interactions. Whilst, to our knowledge, all the bead-rod chain simulations have been carried out with Brownian dynamics simulations~\cite{Liu19895826,Doyle1997251,Petera19997614}, bead-spring chain simulations (which use either FENE~\cite{Aust19995660,Ryder:2006ig} or wormlike chain spring force laws~\cite{Jendrejack20027752,Schroeder20051967}) have treated the solvent either implicitly using Brownian dynamics simulations~\cite{Jendrejack20027752,Schroeder20051967}, or explicitly, using either nonequilibrium molecular dynamics~\cite{Aust19995660} or multiparticle collision dynamics~\cite{Ryder:2006ig} simulations. 

Free-draining bead-rod chain simulations, with chain lengths ranging from 20 beads~\cite{Liu19895826} to 100 beads~\cite{Doyle1997251}, predict a shear thinning exponent of $-(1/2)$ for sufficiently high shear rates. The inclusion of hydrodynamic interactions (for chains with a maximum of 20 beads) appears to decrease the magnitude of the exponent to a value closer to $0.3$~\cite{Petera19997614}. Interestingly, \citet{Liu19895826} observed that either decreasing the number of beads to 2 or using a multibead rod model leads to an exponent equal to $-(1/3)$, which is identical to the analytical prediction by~\citet{Stewart:1972gt} for a rigid dumbbell model. The increase in the magnitude of the predicted power law exponent from (1/3) to (1/2) is consequently purely due to the flexibility of the chain.  

Two different power law regimes of shear thinning are observed in Brownian dynamics simulations of bead-spring chains with finitely extensible springs, with hydrodynamic and excluded volume interactions included~\cite{Jendrejack20027752,Schroeder20051967}. At intermediate Weissenberg numbers, $100 < W\!i < 1000$, the exponent is estimated to be close to $-(1/2)$. However, at high Weissenberg numbers,  \citet{Jendrejack20027752} obtain an exponent equal to $-(2/3)$ (for $W\!i > 10^4$), while~\citet{Schroeder20051967} fit their simulation data with an exponent equal to $-0.61$ for $W\!i > 10^3$. On the other hand, only a single power law regime with an exponent equal to $-0.59\pm 0.02$, is observed in the nonequilibrium molecular dynamics simulations of~\citet{Aust19995660}, who obtain data collapse onto a master plot when they plot their viscosity data (for different chain lengths), in terms of the ratio $\eta_\text{p}/\eta_{\text{p}0}$ versus $\lambda^* \dot \gamma$, where $\lambda^*$ is as defined above. Since the solvent is treated explicitly in these simulations, hydrodynamic interactions are automatically included. The high Weissenberg number asymptotic value of the shear thinning exponent in all these simulations is consistent with analytical results for finitely extensible chains. The multiparticle collision dynamics simulations of~\citet{Ryder:2006ig}, who used a FENE spring force law, were not carried out to sufficiently high Weissenberg numbers in order to observe shear thinning. However, by including a bending potential which enabled them to examine the influence of chain stiffness, they find that with increasing chain stiffness, in contrast to the analytical results of~\citet{Winkler2010}, the viscosity shear thins with an exponent equal to $-(1/3)$, analogous to the prediction for a rigid dumbbell or multibead rod model. 

A striking contrast between predictions of bead-rod and bead-spring chain simulations is that while the former predicts that the viscosity levels off at high shear rates and approaches a second constant Newtonian regime, the latter predicts that the viscosity never levels off, but rather that the power law regime persists indefinitely. These observations are valid regardless of whether hydrodynamic interactions are included or not. However, it should be noted that~\citet{Petera19997614} seem to suggest that even in the case of bead-rod chain simulations, when both hydrodynamic and excluded volume interactions are included, the second Newtonian regime disappears, and the power law regime appears to persist, at least until the maximum value of nondimensional shear rate in their simulations of $\lambda_Z \dot \gamma \approx 10^4$. The existence of the second Newtonian regime in bead-rod simulations has been attributed to a number of different reasons. The free draining bead-rod simulations of~\citet{Liu19895826} show that the mean-square end-to-end distance of a chain increases with increasing shear rate until the onset of the power law regime (where the chain stretch is a maximum), followed by a continuous decrease in magnitude which coincides with the levelling off of the viscosity at high shear rates. By separating the polymer contribution to solution stress into a Brownian and a viscous contribution, \citet{Doyle1997251} demonstrate that power law shear thinning is caused largely by Brownian stresses, while the levelling off of viscosity at high shear rates is attributable to viscous stresses. For finitely extensible bead-spring chains, \citet{Jendrejack20027752} use a simple scaling argument to  explain the observed power law scaling of viscosity with Weissenberg number. They first show that the viscosity scales as $\eta_\text{p} \sim \delta_y^2$, where $\delta_y$ is a measure of chain length fluctuations transverse to the flow direction, and subsequently that, $\delta_y \sim \dot \gamma^ {-1/3}$. Further, they explain the levelling off of viscosity for bead-rod chains by arguing that the predicted scaling of $\delta_y$ does not persist to very high shear rates due the existence of a bond length cutoff at small length scales.

While the rheological behaviour of dilute polymer solutions has been studied extensively over the years, experimental measurements~\cite{Hur2001421,Heo20051117,Heo20088903,Pan2014339,Hsiao:2017fm,Prabhakar:2017bx} and theoretical predictions~\cite{Stoltz2006137,Huang201010107,Jain2012a,Jain2012b,Saadat:2015bg,Prabhakar:2017bx,Sasmal:2017ey} of the rheological behaviour of semidilute polymer solutions are comparatively sparse. ~\citet{Hur2001421} carried out pioneering single molecule and bulk rheological measurements on semidilute solutions of DNA in shear flow. By plotting the polymer contribution to viscosity in the Weissenberg number range, $10< W\!i < 3000$, they find a power law shear thinning behaviour, $\eta_p \sim W\!i^{-\alpha}$, where the exponent has the values, $\alpha = 0.53, 0.43, \text{and} \, 0.51$, for scaled concentrations, $c/c^* = 1, 3, \text{and} \, 6$, respectively, with the Weissenberg number defined in terms of the longest relaxation time, $\lambda_1$, as in the case of dilute solutions discussed above. Noting the similarity in dilute and semidilute solutions of the dependence of the fractional extension of a chain on strain, and the similarity in the time dependence of the power spectral density (when viewed at the same Weissenberg number), along with the similarity of shear thinning exponents, they conclude that the role of the surrounding DNA chains in a semidilute solution is to merely viscosify the solvent. 

More recently, \citet{Heo20088903} have examined the linear and nonlinear rheological behaviour of semidilute solutions of polystyrene in tricresyl phosphate, close to and above the entanglement concentration $c_\text{e}$, with a view to examining if master plots can be constructed by choosing appropriate scaling variables using the ``blob" scaling arguments of de Gennes~\cite{dgen79}. They find that in the concentration range, $c_\text{e} < c < c^{**}$, where $c^{**}$ is the concentration above which a polymer chain obeys random walk statistics on all length scales, data for different molecular weights can be collapsed onto a universal curve when the scaled viscosity, $\eta_\text{p}/\eta_s (c [\eta]_0)^{1/(3\nu -1)}$, is plotted versus $\lambda_\text{e} \dot \gamma$ (in the range $10^{-4} < \lambda_\text{e} \dot \gamma < 30$). Here, $\nu$ is the Flory scaling exponent, and $\lambda_\text{e}$ is the Rouse time of an entanglement strand. Interestingly, curves for different $c/c_\text{e}$ do not collapse on top of each other. \citet{Heo20088903} do not comment on the shear thinning exponent, nor the apparent decrease that is observed in the slope of the scaled viscosity curve at high values of the nondimensional shear rate.

Computing the behaviour of semidilute polymer solutions is challenging since both \emph{intra} and \emph{inter} molecular hydrodynamic and excluded volume interactions need to be taken into account, particularly with the former being a long-range interaction. The seminal bead-spring chain simulations by~\citet{Stoltz2006137} (using Brownian dynamics), and~\citet{Huang201010107} (using multiparticle collision dynamics), were the first attempts to  computationally estimate viscometric functions for semidilute unentangled polymer solutions in simple shear flow. In both these papers, the Weissenberg number is defined  as, $ W\!i_\text{c} = \lambda_1 \dot \gamma$, where $\lambda_1$ is the longest relaxation time obtained by fitting an exponential function to the tail end of the decay of the mean square end-to-end distance of a chain, as it relaxes to equilibrium from an initially extended state. \citet{Huang201010107}  show explicitly from their simulations that $\lambda_1$, which is a function of concentration, obeys the well known scaling law,  $\lambda_1 \sim (c/c^*)^{(2-3\nu)/(3\nu -1)}$, derived from blob scaling arguments. \citet{Stoltz2006137}, in their simulations of chains with 10 beads, do not observe data collapse when they plot the scaled viscosity, $\eta_\text{p}/\eta_s (c /c^*)$, versus $W\!i_\text{c}$  for two concentrations, $c/c^* =1$  and $c/c^* =2$, on the same plot (though this is not entirely clear since error bars have not been reported). By converting their nondimensional simulation results to dimensional quantities (by choosing appropriate parameter values), they show that for $c/c^* =1$, in the Weissenberg number range, $ 3 < W\!i_\text{c} < 200$, a shear thinning exponent equal to $-0.51$  is obtained, in close agreement with the experimental results of~\citet{Hur2001421}. On the other hand, by plotting $\eta_\text{p}/\eta_{\text{p}0}$  versus $W\!i_\text{c}$, \citet{Huang201010107} show that data for different values of $(c /c^*)$ collapse onto a master plot, provided the chain length is held fixed. A weak dependence on chain length is observed when data for chains with 50 and 250 beads are plotted together. For large values of Weissenberg number, regardless of concentration and chain length, they observe power law shear thinning, with an exponent equal to $-0.45$. 

It is clear from the summary of experimental measurements and theoretical predictions of the shear rate dependence of the viscosity of dilute and semidilute polymer solutions given above, that there is a great deal of variation in the reported results. While there are many common elements, there are also significant differences. The aim of the present paper is to gain some understanding of the origin of these differences, and to attempt to resolve them by representing experimental data in terms of nondimensional variables that capture the underlying physics. Towards this end, we are aided by our recent experimental work in characterising the equilibrium and linear viscoelastic behaviour of DNA solutions in terms of the solvent quality $z$ and the scaled concentration $c/c^*$~\cite{Pan2014339,Pan2014b}. Experimental measurements have been carried out on dilute and semidilute unentangled solutions of DNA with three  different molecular weights, at various values of $z$, $c/c^*$, and shear rates $\dot \gamma$. We have also examined the behaviour of synthetic polystyrene solutions. In the case of dilute solutions, we have reinterpreted some of the wealth of prior experimental data on dilute polystyrene solutions in terms of the variables $z$, $c/c^*$ and $\lambda_Z \dot \gamma$. For semidilute solutions, we have carried out experimental measurements for two different molecular weights of polystyrene dissolved in dioctyl phthalate, at various points in the space of coordinates $\{z, c/c^*, \dot \gamma \}$. The characteristics and specific details of all the solutions are discussed in the section below. 

The plan of the paper is as follows. In various subsections of section~\ref{sec:expt},  
we briefly describe the experimental protocol for preparing the DNA and polystyrene samples, and for carrying out the viscosity measurements, with further details deferred to the supplementary material. In section~\ref{zandcstar}, the solvent quality and overlap concentration of the DNA and polystyrene solutions studied here, and of dilute polystyrene solutions examined in prior experimental studies, is tabulated. In the latter case, since this has not been discussed previously in the literature, details of the procedure used to estimate $z$ and $c^*$ are given in Appendix~\ref{sec:HuaWu}. The shear rate dependence of dilute    
solutions is considered in section~\ref{sec:dilDNAshear}, while that of semidilute solutions is considered in section~\ref{sec:sdshflow}, including the estimation of relaxation times and zero shear rate viscosities. A detailed comparison of the master plots that describe the shear thinning behaviour of dilute DNA and polystyrene solutions is carried out in section~\ref{sec:dilshear}. In subsection~\ref{sec:semidilshear} it is shown that the use of a conventionally defined relaxation time to construct master plots for semidilute solutions does not lead to data collapse. A scaling argument based on the formation of Pincus blobs~\cite{Pincus1976386} in shear flow is used in section~\ref{subsec:colbyrelax} to derive an alternative relaxation time which has a different dependence on $c/c^*$. Some simple scaling expressions that have been derived previously for semidilute solutions at equilibrium are given in Appendix~\ref{blob} for ease of reference, while the existence of Pincus blobs in semidilute DNA solutions subjected to shear flow (at typically measured shear rates), is discussed in Appendix~\ref{pincblob}. The usefulness of the alternative relaxation time in revealing the universal behaviour of semidilute solutions is examined in subsection~\ref{subsec:rouse}. Finally, in section~\ref{sec:conc}, we summarize the principal conclusions of this work.

\begingroup
\small
\begin{table}[t]
  \caption{\label{tab:DNAprop} Representative properties of the 25 kbp, $\lambda$-phage, and T4 DNA
    used in this work (reproduced from Table I of Ref.~\citenum{Pan2014339}). The contour length is estimated using the
    expression $L$ = number of base-pairs $\times \, 0.34$ nm; the
    molecular weight is calculated from $M$ = number of base-pairs
    $\times \, 662$ g/mol (where the base-pair molecular weight has
    been calculated for a sodium-salt of a typical DNA base-pair segment);
    the number of Kuhn steps from 
    $N_{\mathrm{k}}$ = $L / (2 P)$ (where $P$ is the persistence
    length, which is taken to be 50 nm), and the radius of gyration at
    the $\theta$ temperature is estimated from $\Rgth\ = L/\sqrt{6
      N_{\mathrm{k}}}$. }
\vskip10pt
\setlength{\tabcolsep}{10pt}
{\def\arraystretch{1.2}
\begin{tabular}{ c  c  c  c  c }
\hline
DNA Size       & $M$ 
            & $L$
            & $N_{\mathrm{k}}$
            & $R_{\mathrm{g}}^{\theta}$
\\
(kbp)
            & ($\times 10^{6}$ g/mol)
            &($ \mu$m)
            & 
            & (nm)
\\
\hline
\hline
25
            & 16.6
            & 9
            & 85
            & 376
\\
48.5
            & 32.1
            & 16
            & 165
            & 524     
\\
165.6
            & 110
            & 56
            & 563
            & 969
\\
\hline
\end{tabular}
}
\end{table}
\endgroup

\section{\label{sec:expt} Materials and method}
\subsection{DNA samples}

Three different linear genomic DNA samples, (i) 25 kilobasepairs (kbp), (ii) $\lambda$-phage  [48.5 kbp] and (iii) T4 bacteriophage [165.6 kbp], were used in this work at various temperatures (10 to 44.6\degC), and at concentrations in the dilute (0.112 to 0.0038 mg/ml) 
and semidilute (0.441 to 0.023 mg/ml) regimes. The 25 kbp samples were extracted, linearized, and purified from \emph{Escherichia coli} (\emph{E. coli}) stab cultures procured from Smith's group at UCSD, who have genetically engineered special 3-300 kbp double-stranded DNA fragments. These fragments have been incorporated inside \emph{E. coli} bacterial strains, which can be cultured to produce sufficient replicas of its DNA, which are then cut at precisely desired locations to extract the  fragments~\cite{Laib20064115}. The linear genomic DNA of $\lambda$-phage was purchased from New England Biolabs, U.K. (\#N3011L), and the linear genomic DNA of T4 phage was purchased from Nippon Gene, Japan (\#314-03973). 

Procedures for preparation and quantification of the DNA samples, details of the solvent used, and estimation of DNA concentration, etc., along with the typical properties of the DNA molecules have been reported in detail in our earlier work~\cite{Pan2014339,Pan2014b}. For ease of reference, a summary of the key aspects are reproduced here in the supplementary material. The molecular weight, contour length, number of Kuhn steps, and the radius of gyration at $\theta$-temperature, of the DNA molecules used in this work, are given in Table~\ref{tab:DNAprop}. They have been reproduced here from a similar Table in \citet{Pan2014339}.

\subsection{Polystyrene samples}

As discussed in the introduction, two different approaches have been adopted here with regard to polystyrene samples. In the case of dilute solutions, we have adapted the previously published results of Inagaki and coworkers~\cite{Kotaka:1966cl,Suzuki:1969ec}, \citet{Noda:1968gt} and~\citet{Hua2006787}, since it is possible to convert the data reported in their work to the framework adopted in this paper, namely, representation of the results in terms of the non-dimensional variables $(z, c/c^*, \lambda_Z \dot \gamma)$. The conversion procedure is discussed in detail in Appendix~\ref{sec:HuaWu}, with some of the relevant parameters reproduced in this section,  wherever relevant. Details of polymer sample sources, the polydispersity index, and sample preparation procedure have been given in the original references, to which we refer the reader. 

In the semidilute regime, we have carried out experiments on two linear polystyrene polymers: (i) with molecular weight 1.14 $\times$ 10$^{6}$ g/mol (1.1 M) and a polydispersity index (PDI) = 1.09 purchased from Polymer Source Inc. (Canada) and (ii) with molecular weight
1.54 $\times$ 10$^{7}$ g/mol (15.4 M) (PDI = 1.04) purchased from
Varian (England). Both the polystyrene samples were dissolved in DOP, which is considered a $\theta$-solvent for polystyrene at 22\degC~\citep{Brandrup1999}. To assist with the dissolving of polystyrene in DOP, methylene chloride was used as a co-solvent and the mixture was mixed for 24 hours. Methylene chloride was then completely evaporated in a vacuum oven at 40\degC\ over the course of several days until no further weight loss was registered.

\subsection{Shear rheometry}

A Contraves Low Shear 30 rheometer (1T/1T -- cup and bob; shear rate $\dot{\gamma}$ range: 0.01--100 $s^{-1}$; temperature sensitivity: $\pm$ 0.1\degC) has been used to obtain all the shear viscosity measurements reported in the present work. The Contraves Low Shear 30 rheometer has two main advantages: (i) it has a zero shear rate viscosity sensitivity even at a shear rate of 0.017 $s^{-1}$ and thus can measure very low viscosities; and (ii) has a very small sample requirement (minimum 800 $\mu$l)~\cite{Heo20051117}. These advantages are ideal for measuring the viscosities of DNA solutions. Details of the measurement protocol followed here are given in the supplementary material.

\section{\label{zandcstar} Solvent quality and overlap concentration}

The state of a polymer solution at equilibrium is completely determined by the solvent quality parameter $z$, defined by the expression~\citep{Schafer99},
 \begin{equation}
 \label{eq:z}
z = k\left(1- \dfrac{T_{\theta}}{T} \right) \, \sqrt{M}
\end{equation}
(where $k$ is a chemistry dependent constant, and $T_{\theta}$ is the $\theta$-temperature), and the overlap concentration $c^*$, which is defined by the expression~\cite{RubCol03},
 \begin{equation}
 \label{eq:c*}
 c^{*} = \frac{M }{ \left( \dfrac{4 \pi}{3} \right) \, \Rg^{3} \,N_{A} }
 \end{equation}
(where \Rg\ is the radius of gyration). The values of $z$ and $c^*$ for all the solutions for which experimental measurements of viscosity have been reported in this work are tabulated below, firstly for DNA solutions, and subsequently for the various sets of polystyrene solutions. 

\subsection{\label{subsec:DNAzc*} DNA solutions}

\begin{table}[tb]
\caption{Solvent quality parameter $z$ and overlap concentration $c^{*}$ (in mg/ml) for 25 kbp, $\lambda$-phage, and T4 DNA, at various temperatures. The $\theta$-temperature is taken to be 15\degC. } 
\vskip10pt
\centering 
\setlength{\tabcolsep}{4pt}
{\def\arraystretch{1.1}
\begin{tabular}{l  c c c ccccc} 
\hline 
 &  &  & & 15\degC & 20\degC  & 25\degC & 30\degC & 35\degC \\ 
\hline \hline
&  & $z$ & & 0 & 0.33 & 0.64 & 0.95 &1.24  \\
\raisebox{1.5ex}{25 kbp}  &  & $c^{*}$ &  & 0.123 & 0.084 & 0.068 & 0.059 & 0.052  \\
\hline
&  & $z$ & & 0 & 0.45 & 0.89 & 1.32 & 1.73  \\
\raisebox{1.5ex}{48.5 kbp}  &  & $c^{*}$ &  & 0.088 & 0.055 & 0.043 & 0.037 & 0.032  \\
\hline
&  & $z$ & & 0 & 0.84 & 1.65 & 2.44 & 3.20  \\
\raisebox{1.5ex}{165.6 kbp}  &  & $c^{*}$ &  & 0.048 & 0.024 & 0.018 & 0.015 & 0.013  \\
\hline 
\end{tabular}
}
\label{tab:zc}
\end{table}

The value of $z$ at any temperature and molecular weight can be determined from Eq.~(\ref{eq:z}) if the values of $k$ and $T_{\theta}$ are known. The procedure for determining the value of  $k= 0.0047 \pm 0.0003 \, \text{(g/mol)}^{-1/2}$, and $T_{\theta} =15$ \degC, for the DNA solutions used in this study, has been discussed extensively in Ref.~\cite{Pan2014339}. Typical values of $z$ for the three DNA samples used here, at various values of $T$, are reported in Table~\ref{tab:zc}.

The radius of gyration can be determined from the expression $\Rg = \Rgth \, \ag (z)$, for any $M$ and $T$, where $\ag (z)$ is the universal swelling of the radius of gyration for dilute polymer solutions~\cite{Schafer99}. Since the chain conformations at the $\theta$ temperature are expected to be ideal Gaussian chains, the analytical value for the radius of gyration at $T_{\theta}$ is, $\Rgth = L/\sqrt{6 N_{\mathrm{k}}}$. We have consequently used the respective values of $L$ and $N_{\mathrm{k}}$ for all the molecular weights used here, to determine \Rgth\ (as displayed in Table~\ref{tab:DNAprop}). Further, since we know $z$, \ag\ can be determined from the expression $\alpha_{\mathrm{g}} = (1 + a\, z + b\, z^{2} + c \, z^{3})^{m/2}$, where the constants, $a = 9.528$, $b = 19.48$, $c = 14.92$, and $m = 0.1339$ have been obtained by Brownian dynamics simulations~\citep{Kumar20037842}. Representative values of $c^{*}$ found using this procedure, at various $M$ and $T$, are displayed in Table~\ref{tab:zc}. The validity of the values of $\Rg$ (and consequently $c^*$) determined by this procedure has been discussed by~\citet{Pan2014339}.

\subsection{Polystyrene solutions}

\subsubsection{\label{ultra} Ultradilute solutions in the limit ${c/c^* \to 0}$}

\begin{table}[tb]
\caption{\label{tab:Noda} Solvents, polystyrene molecular weights, contour lengths, number of Kuhn steps, temperatures  and solvent qualities of the samples used in the experiments of Inagaki and coworkers~\cite{Kotaka:1966cl,Suzuki:1969ec} (superfix $a$) and \citet{Noda:1968gt} (superfix $b$). In the latter work, a mixture of \textit{cis} and \textit{trans} isomers of the solvent decalin has been used, with 60\% of the  \textit{cis} isomer. Note the difference in the $\theta$-temperature reported for decalin in the two works. The number of Kuhn steps is estimated based on the molar mass of a Kuhn monomer for polystyrene (720 g/mol), and the contour length is calculated from the length of a single Kuhn step (1.8 nm)~\cite{RubCol03}. }
\vskip5pt
\setlength{\tabcolsep}{2pt}
{\def\arraystretch{1.2}
{\begin{tabular}{ l  c  c  c c c}
\hline
\multirow{2}{*}{Solvent}     & $M$  &  $L$ & $N_\text{k}$ & $T$  & {$z$} \\  
     &  ($\times 10^{6}$ g/mol) & ($\mu$m) & & (\degC) &   \\
\hline
\hline
\multirow{3}{*}{Benzene$^a$} & 7.14 & 17.85 & 9917 & \multirow{3}{*}{30} & 13.6 \\
& 3.16 & 7.90 &4389 & & 9.0 \\
& 1.39 & 3.48 & 1931 & & 6.0 \\
\hline
\multirow{2}{*}{1-Chlorobutane$^a$} & 7.14 & -- & -- & \multirow{2}{*}{38} & 4.7 \\
& 3.16 & --  &  --  & &3.1 \\
\hline
\textit{Trans}-decalin$^a$ & 7.14 &  --  &  --   & 23.8 ($\theta$) & 0 \\
Toluene$^b$ & 13.6 & 34 &  18889 & 25 & 5.1 \\
\textit{Mix}-decalin$^b$ & 13.6 &  --  &   --  & 15.4 ($\theta$) & 0 \\
\hline
\end{tabular}
}}
\end{table}

Inagaki and coworkers~\cite{Kotaka:1966cl,Suzuki:1969ec} and \citet{Noda:1968gt} have carried out viscosity measurements on a series of dilute polystyrene solutions. Using Zimm-Crothers and Ubbelohde viscometers, they have estimated the intrinsic viscosity $\ivisc$ by extrapolating finite concentration data to the limit of zero concentration. With this procedure, they have tabulated measured values of the ratio $\ivisc/\ivisc_0$ as a function of $\lambda_Z  \dot \gamma$. (This data is discussed further in section~\ref{sec:dilshear} below). For the present it suffices to note that since measurements are reported in the ultradilute limit, $c/c^* \to 0$, it is not necessary to estimate $c^*$ for these solutions. 

The solvent quality of the solutions used in Refs.~\cite{Kotaka:1966cl}, \cite{Noda:1968gt}, and \cite{Suzuki:1969ec}  can be determined from values of the swelling of the intrinsic viscosity, $\alpha_\eta =  \left( [\eta]_0/ [\eta]_{0,\theta} \right)^{1/3}$, which have been reported for most of the samples. The universal swelling of the viscosity radius $\alpha_{\eta}$, as a function of the solvent quality $z$, has been determined by Brownian dynamics simulations for dilute polymer solutions~\citep{Pan2014b}, and the simulation data has been shown to satify the expression $\alpha_\eta = (1 + a\, z + b\, z^{2} + c \, z^{3})^{m/2}$, where the values of the constants are: $a = 5.448$, $b = 3.156$, $c = 3.536$, and $m = 0.1339$. It is consequently straightforward to invert the function $\alpha_\eta (z)$ to find the values of the solvent quality $z$ that correspond to the measured values of $\alpha_\eta$. The solvents, polystyrene molecular weights, experimental temperatures, and the estimated solvent qualities are given in Table~\ref{tab:Noda}. Note that for the molecular weights $1.39 \times 10^{6}$ g/mol and $3.16 \times 10^{6}$ g/mol, in the solvents Benzene and 1-Chlorobutane, respectively, where the swelling $\alpha_\eta$ has not been reported by~\citet{Suzuki:1969ec}, it is straightforward to calculate the solvent quality from $z_1 = z_2 \sqrt{M_1/M_2}$, where $z_2$ is the known solvent quality at molecular weight $M_2$, at the same temperature $T$. 

\begin{table}[t]
\caption{\label{tab:zcPS} Molecular weights, contour lengths, number of Kuhn steps, solvent quality and  overlap concentration (in g/ml) for dilute polystyrene samples PS1 to PS4 used in Ref.~\cite{Hua2006787}, at various temperatures $T$. The contour length and the number of Kuhn steps are calculated as described in the caption to Table~\ref{tab:Noda}. } 
\vskip10pt
\centering 
\setlength{\tabcolsep}{1.8pt}
{\def\arraystretch{1.05}
\begin{tabular}{l  c c c c c c c c c c} 
\hline 
&  $M$ &  $L$ & $N_\text{k}$ & & & & \multicolumn{4}{c}{$T$  (\degC) }  \\ 
\cline{8-11}
 &  ($\times 10^{6}$ g/mol) & ($\mu$m) & & & & & 22 & 25  & 35 & 45  \\ 
\hline \hline
&  & & & & $z$ & & 0 & 0.041 & 0.17 & 0.30   \\
\raisebox{1.5ex}{PS1 }  & \raisebox{1.5ex}{0.55} & \raisebox{1.5ex}{1.38}& \raisebox{1.5ex}{764}&  & $c^{*}$ &  & 0.021 & 0.019 & 0.016 & 0.014   \\
\hline
&  & & & & $z$ & & 0 & 0.046 & 0.19 & 0.33   \\
\raisebox{1.5ex}{PS2 }  & \raisebox{1.5ex}{0.68} & \raisebox{1.5ex}{1.70}& \raisebox{1.5ex}{944}& & $c^{*}$ &  & 0.018 & 0.017 & 0.014 & 0.012   \\
\hline
&  & & & & $z$ & & 0 & 0.053 & 0.22 & 0.38   \\
\raisebox{1.5ex}{PS3 }  &\raisebox{1.5ex}{0.93} & \raisebox{1.5ex}{2.33}& \raisebox{1.5ex}{1292} &  & $c^{*}$ &  & 0.016 & 0.014 & 0.012 & 0.010   \\
\hline 
&  & & & & $z$ & & 0 & 0.078 & 0.33 & 0.56   \\
\raisebox{1.5ex}{PS4 } & \raisebox{1.5ex}{2} &  \raisebox{1.5ex}{5.0}& \raisebox{1.5ex}{2778} &  & $c^{*}$ &  & 0.011 & 0.009 & 0.007 & 0.006   \\
\hline 
\end{tabular}
}
\end{table}

\subsubsection{Dilute solutions  at finite concentrations below ${c^*}$}

\citet{Hua2006787} have reported viscosity measurements for four different polystyrene molecular weight samples dissolved in dioctyl phthalate (DOP), with a Couette rheometer at the $\theta$-temperature (22\degC), and three other temperatures in the good solvent regime (25\degC, 35\degC\ and 45\degC). In order to represent the measurements of~\citet{Hua2006787} in terms of the non-dimensional variables $(z, c/c^*, \lambda_Z \dot \gamma)$, a number of steps are necessary, centered around the estimation of intrinsic viscosities $\ivisc_{0}$ at all the reported temperatures. The detailed procedure is discussed in Appendix~\ref{sec:HuaWu}. Here, the values of $z$ and $c^*$ estimated in this manner are tabulated in Table~\ref{tab:zcPS}, where the nomenclature used by them has been adopted.

\subsubsection{Solutions  at finite concentrations above ${c^*}$}

\begin{table}[tb]
\caption{Molecular weights, contour lengths, number of Kuhn steps, solvent quality parameter and overlap concentration (in mg/ml) for polystyrene solutions PS1.1M and PS15.4M, used to carry out viscosity measurements in the semidilute regime, at various temperatures. The $\theta$-temperature is taken to be 22\degC. The contour length and the number of Kuhn steps  are calculated as described in the caption to Table~\ref{tab:Noda}. } 
\vskip5pt
\centering 
\setlength{\tabcolsep}{1.5pt}
{\def\arraystretch{1.1}
\begin{tabular}{l  c c c c c c c c c} 
\hline 
&  $M$ &  $L$ & $N_\text{k}$ & & &  &  \multicolumn{3}{c}{$T$  (\degC) }  \\ 
\cline{8-10}
 &   ($\times 10^{6}$ g/mol) &  ($\mu$m) &  &  & &  & 22 & 30  & 35 \\ 
\hline \hline
& & & &  & $z$ & & 0 & 0.155 & 0.248   \\
\raisebox{1.5ex}{PS1.1}  & \raisebox{1.5ex}{1.1} & \raisebox{1.5ex}{2.75} &  \raisebox{1.5ex}{1528 }& & $c^{*}$ &  & 0.014 & 0.011 & 0.010   \\
\hline
& & & &  & $z$ & & 0 & 0.570 & 0.911   \\
\raisebox{1.5ex}{PS15.4}  & \raisebox{1.5ex}{15.4}  & \raisebox{1.5ex}{38.5} & \raisebox{1.5ex}{21389 } & & $c^{*}$ &  & 0.0036 & 0.0021 & 0.0018   \\
\hline
\end{tabular}
}
\label{tab:PSzc}
\end{table}

It is necessary to evaluate the solvent quality and the overlap concentration for the two linear polystyrene polymers with molecular weights 1.14 $\times$ 10$^{6}$ g/mol (PS1.1M) and 1.54 $\times$ 10$^{7}$ g/mol (PS15.4M), which have been used for the experiments in the semidilute regime, in order to locate the samples in the space of coordinates $(z, c/c^*, \lambda \dot \gamma)$. These properties are evaluated from the properties of the solution in the dilute regime. The solvent quality $z$ for these solutions can be calculated using the value of $k$ for polystyrene/DOP solutions determined as discussed in Appendix~\ref{sec:HuaWu}. Since the dependence on molecular weight of the intrinsic viscosity at the theta temperature is known from fitting the data reported in Ref.~\cite{Hua2006787} ($\ivisc_{0,\theta} \sim M^{0.52}$), the values of  $\ivisc_{0,\theta}$ for the PS1.1M and PS15.4M solutions can be determined. Using the procedure described in Appendix~\ref{sec:HuaWu}, it follows that the values of $\ivisc_0 (z)$,  and consequently $c^*(z)$ can be determined at the three temperatures, 22, 30, and $35 \degC$,  at which the viscosity measurements have been carried out. The values of $z$ and $c^*$ found in this manner are listed in Table~\ref{tab:PSzc}.

\section{\label{sec:dilDNAshear} Dilute solutions}
\subsection{Relaxation time $\pmb{\lambda_Z}$ and zero shear rate viscosity $\pmb{\eta_{\text{p}0}}$}

For the purpose of examining the universality of the viscosity versus shear rate data for dilute polymer solutions, we need the relaxation time $\lambda_Z$ in order to non-dimensionalize the shear rate, and the zero shear rate viscosity $\eta_{\text{p}0}$  in order to non-dimensionalize the polymer contribution to viscosity. The values of these two quantities, for the various solutions considered here, are discussed below before assessing the existence of master plots.

\subsubsection{\label{subsec:DNAlambda} DNA solutions}

\begin{table}[tb]
\addtolength{\tabcolsep}{2pt}
\caption{\label{tab:ivisc} Intrinsic viscosities $\ivisc_0$ (in ml/mg) and relaxation times $\lambda_Z$ (in s), for 25 kbp, $\lambda$-phage DNA, and T4 DNA, at various temperatures $T$. The relaxation time is calculated using Eq.~(\ref{eq:lambdaZ}).}
\vspace{2pt}
\centering
\setlength{\tabcolsep}{4pt}
{\def\arraystretch{1.1}
\begin{tabular}{l c c c c c c c c}
\hline
$T$ & \multicolumn{2}{c}{25 kbp} &  & \multicolumn{2}{c}{$\lambda$-phage DNA} &  & \multicolumn{2}{c}{T4 DNA} \\
\cline{2-3} \cline{5-6}  \cline{8-9}
 (\degC) & $\ivisc_0$ & $\lambda_Z$ &  & $\ivisc_0$ &$\lambda_Z$ &  & $\ivisc_0$ & $\lambda_Z$ \\ 
\hline
\hline
15  & 7.5  & 0.059 & & 16.5 & 0.252 & & 28.7  & 1.50  \\

25 & 9.8  & 0.058 & & 26.6 & 0.307 &  & 56.9  & 2.24  \\

30 & 12.8  & 0.067 & & 30.6 & 0.311 &  & 69.2  & 2.40 \\

35 & 13.8 & 0.064 & & 34.3 & 0.310 &  & 77.3 & 2.38 \\
\hline
\end{tabular}
}
\end{table}

By appropriately extrapolating concentration dependent functions of the zero shear rate viscosity to zero concentration, \citet{Pan2014b} have previously determined the intrinsic viscosity $\ivisc_0$ for the 25 kbp and T4 DNA solutions used here. For $\lambda$-phage DNA (which is not discussed in Ref.~\cite{Pan2014b}), the intrinsic viscosity $[\eta]_{0,\theta}$ at the $\theta$-temperature \Ttheta\ was determined here by following the same procedure. The value of $\ivisc_0$ for $\lambda$-phage DNA  at all other temperatures has been calculated using the expression $\ivisc_0 = [\eta]_{0,\theta} \, [\alpha_{\eta}(z) ]^{3}$, with $\alpha_{\eta} (z)$ being the universal swelling of the viscosity radius obtained by Brownian dynamics simulations~\citep{Pan2014b}, discussed in section~\ref{ultra} above. It was established in Ref.~\cite{Pan2014b} that the swelling of the viscosity radius of DNA solutions in the presence of excess salt can be collapsed onto universal master plots determined previously for synthetic polymer solutions. The values of $\ivisc_0$ obtained by this procedure at different temperatures, for all three samples of DNA, are listed in Table~\ref{tab:ivisc}.

The relaxation times $\lambda_Z$, which can be easily determined from Eq.~(\ref{eq:lambdaZ}) using the known values of $\ivisc_0$ for all the three DNA solutions, are also listed in Table~\ref{tab:ivisc}, at the various experimentally relevant temperatures. 

The zero shear rate solution viscosity, $\eta_0$, for dilute solutions of 25 kbp and T4 DNA was determined by us previously by least-square fitting the values of viscosity in the plateau region of very low shear rates (across a range of temperatures and concentrations below $c^*$), and extrapolating to zero shear rate~\citep{Pan2014b}. Values of $\eta_0$ for $\lambda$-phage DNA have been determined here by following the same procedure. Data for the all the three DNA samples, for various values of $c/c^*$ and $T$, are displayed in Table~I of the supplementary material.

\begin{table}[t]
\addtolength{\tabcolsep}{2pt}
\caption{\label{tab:iviscPS} Intrinsic viscosities $\ivisc_0$ (in ml/mg) and relaxation times $\lambda_Z$ (in s) for the polystyrene solutions used in Ref.~\cite{Hua2006787}.}
\vspace{2pt}
\centering
\begin{small}
\setlength{\tabcolsep}{2pt}
{\def\arraystretch{1.1}
\begin{tabular}{l c c c c c c c c c c c}
\hline
$T$ & \multicolumn{2}{c}{PS1} &  & \multicolumn{2}{c}{PS2} &  & \multicolumn{2}{c}{PS3}  &  & \multicolumn{2}{c}{PS4} \\
\cline{2-3} \cline{5-6}  \cline{8-9} \cline{11-12}
 (\degC) & $\ivisc_0$ & $\lambda_Z$ &  & $\ivisc_0$ &$\lambda_Z$ &  & $\ivisc_0$ & $\lambda_Z$ &  & $\ivisc_0$ & $\lambda_Z$ \\ 
\hline
\hline
22  & 0.072  & 0.0011 & & 0.082 & 0.0015 & & 0.096  & 0.0024  & & 0.142  & 0.0076\\

25 & 0.075  & 0.0009 & & 0.086 & 0.0013 &  & 0.101  & 0.0021  & & 0.153  & 0.0067\\

35 & 0.083  & 0.0006 & & 0.096 & 0.0008 &  & 0.115  & 0.0013 & & 0.180  & 0.0045\\

45 & 0.090 & 0.0004 & & 0.104 & 0.0005 &  & 0.125 & 0.0009 & & 0.201  & 0.0031 \\
\hline
\end{tabular}
}
\end{small}
\end{table}

\subsubsection{\label{subsec:PSlambda} Polystyrene solutions}

Since Inagaki and coworkers~\cite{Kotaka:1966cl,Suzuki:1969ec} and \citet{Noda:1968gt} have directly listed values of the ratio $\ivisc/\ivisc_0$ as a function of $\lambda_Z  \dot \gamma$ in their papers, it is not necessary to evaluate $\lambda_Z$ or $\eta_{\text{p}0}$ for their polystyrene solutions. On the other hand, as described in Appendix~\ref{sec:HuaWu}, the data presented by~\citet{Hua2006787} can be used to determine the values of $\ivisc_0$ and $\lambda_Z$ for the solutions used in their work. These values are reproduced in Table~\ref{tab:iviscPS} for ease of reference, and for comparison with the values for DNA solutions in Table~\ref{tab:ivisc}.

\begin{figure}[tbp]
\centering
\resizebox{8.5cm}{!} {\includegraphics*{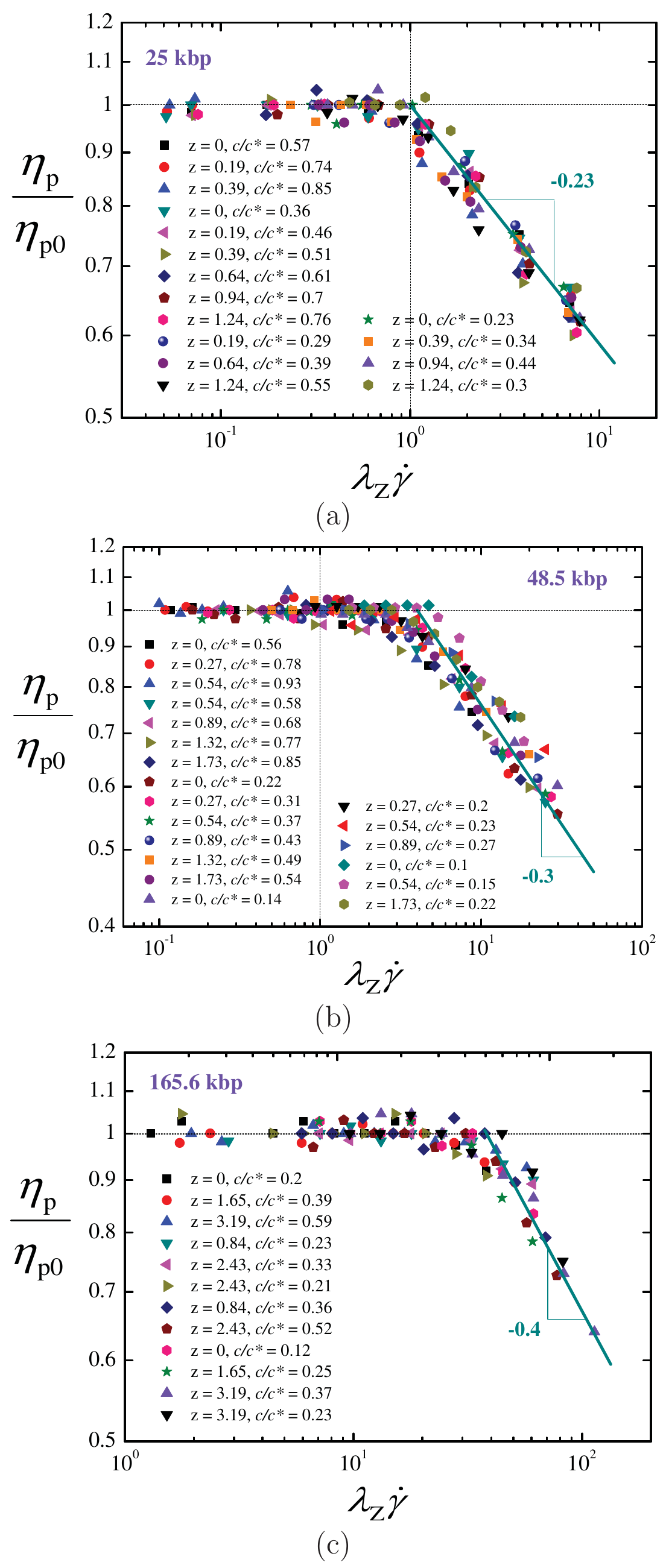}}
\vspace{-10pt}
\caption{Shear rate dependence of the normalised polymer contribution to
viscosity for all the three dilute DNA solutions. When represented in terms of
a relaxation time based on the intrinsic viscosity ($\lambda_{\mathrm{Z}}$),
data for different scaled concentrations (\ccs) and solvent qualities ($z$)
collapse on a master curve for a particular molecular weight in the dilute regime.
The $z$ and corresponding \ccs\ values have been indicated against each
symbol. The terminal slope in each case was obtained by a least-squares
fit to the data at high shear rates.}
\label{fig:diluteshearDNA}
\end{figure}

\begin{figure}[tbp]
\centering
\resizebox{8.5cm}{!} {\includegraphics*{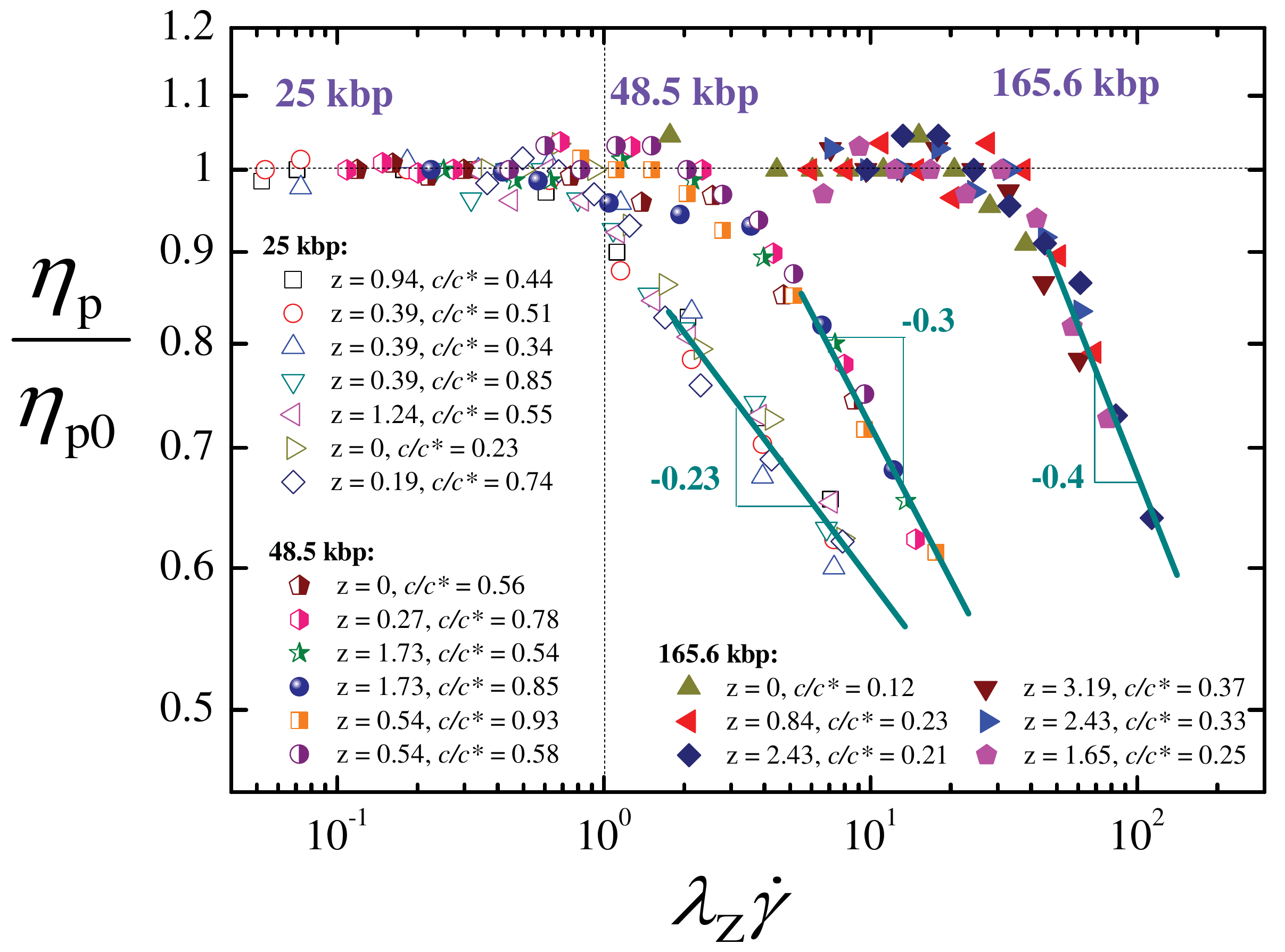}}
\vspace{-5pt}
\caption{Representative data from Figs.~\ref{fig:diluteshearDNA}~(a) to~(c) 
for dilute DNA solutions in shear flow. The data for 25, $\lambda$-phage 
and T4 DNA are represented by open, half-filled and filled symbols,
respectively. The onset of shear thinning shows a molecular weight dependence.}
\label{fig:diluteshearallDNA}
\end{figure}

The steady state solution shear viscosity, $\eta$, was measured for the polystyrene solutions PS1 to PS4 by~\citet{Hua2006787} across a range of temperatures. As in the case of DNA solutions, we have extracted the zero shear rate viscosities $\eta_0$  at each temperature by least-square fitting the values of viscosity in the plateau region of very low shear rates, and extrapolating to zero shear rate. By estimating the solvent viscosity, $\eta_s$, for dioctyl phthalate solutions at various temperatures (as described in Appendix~\ref{sec:HuaWu}), values of the polymer contribution to steady state zero shear rate viscosity, $\etapo$, have been obtained for all the polystyrene solutions used by~\citet{Hua2006787}. Data obtained in this manner, for various values of $c/c^*$ and $T$, are displayed in Table~II of the supplementary material.

\begin{figure}[!bt]
\centering
\resizebox{8.5cm}{!} {\includegraphics*{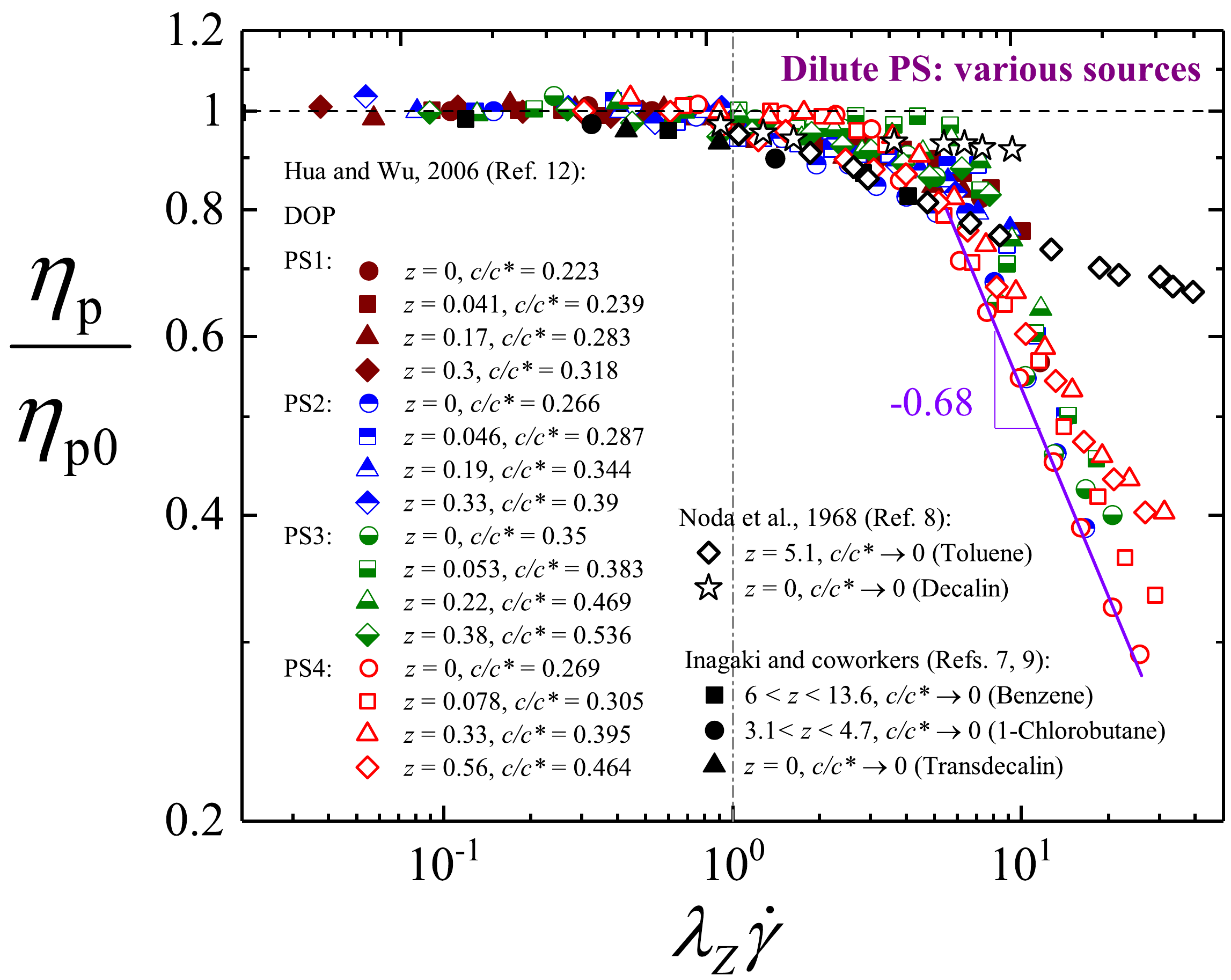}}
\vspace{-5pt}
\caption{Shear thinning of dilute polystyrene solutions. Data reproduced from Refs.~\cite{Kotaka:1966cl}, \cite{Noda:1968gt}, \cite{Suzuki:1969ec}, and~\cite{Hua2006787}, in terms of the variables $z$ and $c/c^*$}
\label{fig:diluteshearPS}
\end{figure}

\subsection{\label{sec:dilshear} Dependence of viscosity on shear rate}

The steady state solution viscosity $\eta$ of the three DNA solutions was measured as a function of shear rate $\dot{\gamma}$ for a variety of concentrations in the dilute regime, and across a temperature range in the crossover regime from $\theta$ to good solvents (15--35\degC). The dependence of $\eta$ on $\dot{\gamma}$ at fixed concentration and at various temperatures, and at fixed temperature and various concentrations, is displayed in Figs.~3~(a)~to~3~(f) of the supplementary material, along with a discussion of the observed behaviour. The data presented in these figures is also given in tabular form in Tables~V, VI, and~VII of the supplementary material, so that they are readily available for comparison with future model predictions for the shear thinning of dilute polymer solutions. 

Here we focus on the behaviour of the scaled viscosity $\eta_\text{p}/\eta_{\text{p}0}$ as a function of non-dimensional shear rate $\lambda_Z \dot \gamma$, as displayed in Figs.~\ref{fig:diluteshearDNA}~(a) to~(c), where the `raw' $\eta$ versus $\dot{\gamma}$ data displayed in Figs.~3 of the supplementary material, has been re-plotted, along with some additional measurements, in terms of these variables. The temperature dependence is represented in terms of the solvent quality $z$, and the concentration dependence in terms of the scaled concentration $c/c^*$. Though there is a degree of scatter, the data appears to collapse onto master curves, independent of solvent quality and concentration. The collapse of data for different temperatures is similar to earlier observations by~\citet{Kotaka:1966cl} and~\citet{Suzuki:1969ec}. In particular, the large difference between the viscosity curves in $\theta$ and good solvents observed by~\citet{Noda:1968gt}, is not reflected in Figs.~\ref{fig:diluteshearDNA}~(a) to~(c). 

Several aspects of the data displayed in Figs.~\ref{fig:diluteshearDNA}~(a) to~(c) are worth noting in the context of previous observations of shear thinning in dilute polymer solutions. First, the magnitude of the shear thinning exponent in all cases is less than $(2/3)$, which is the known limiting value for finitely extensible chains. Second, none of the curves level off at the highest shear rates measured here. Finally, the onset of shear thinning occurs at higher values of $\lambda_Z \dot \gamma$ with increasing molecular weight of the DNA chains. This last observation is more clearly displayed in Fig.~\ref{fig:diluteshearallDNA}, where selected data for all the three DNA samples is plotted together in a single graph. The increase, with increasing molecular weight, in the value of $\lambda_Z \dot \gamma$ at which the viscosity begins to decrease, is in line with the earlier observations of~\citet{Noda:1968gt} for polystyrene in decalin and poly($\alpha$-methyl-styrene). Since the largest nondimensional shear rates measured here are of $\mathcal{O}(10^2)$, the data in Figs.~\ref{fig:diluteshearDNA}~(a) to~(c) is probably still in the crossover region, and the shear thinning exponent may approach the limiting value of  $-(2/3)$ only at much higher shear rates. In the crossover regime, the slope of the viscosity curve is expected to vary from 0 in the Newtonian regime to $-(2/3)$ in the finite extensibility limit. The difference in the power law slopes for the three DNA samples is probably due to their being determined over different ranges of shear rates in the crossover regime. All these aspects are discussed in greater detail after considering the data displayed in Fig.~\ref{fig:diluteshearPS}.

Figure~\ref{fig:diluteshearPS} displays the scaled viscosity versus nondimensional shear rate measurements for dilute polystyrene solutions, reported previously in Refs.~\cite{Kotaka:1966cl}, \cite{Noda:1968gt}, \cite{Suzuki:1969ec}, and~\cite{Hua2006787}, in terms of the variables $z$ and $c/c^*$. It is clear that the onset of shear thinning occurs for shear rates in the range $1 \le \lambda_Z \dot \gamma \le 2$, independent of the molecular weight. Though there is a degree of scatter in the data (roughly of order 5\% about the mean), there is considerable overlap in the nondimensional viscosity values for  $\lambda_Z \dot \gamma \lesssim \mathcal{O}(2)$, independent of solvent quality. All the data of~\citet{Kotaka:1966cl} and~\citet{Suzuki:1969ec}, and the data of~\citet{Noda:1968gt} for the particular case of polystyrene in toluene, seem to agree with that of~\citet{Hua2006787} up to $\lambda_Z \dot \gamma \sim \mathcal{O}(3)$. However, for polystyrene in decalin (which is a $\theta$-solvent), the data of~\citet{Noda:1968gt} seems to suggest that shear thinning stops beyond $\lambda_Z \dot \gamma \sim \mathcal{O}(2)$, and the viscosity levels off at higher shear rates. There is significantly more variation in the behaviour of polystyrene samples whose viscosity was measured at nondimensional shear rates greater than $\lambda_Z \dot \gamma \sim \mathcal{O}(3)$. Firstly, the shear thinning of the 13.6 million molecular weight sample of polystyrene in toluene~\cite{Noda:1968gt} appears to be considerably less than that of all the polystyrene samples considered by~\citet{Hua2006787}, with the viscosity levelling off at high shear rates. Some levelling off can also be seen in the data of~\citet{Hua2006787} for the 2 million molecular weight samples with increasing solvent quality. At the $\theta$-temperature, the power law shear thinning exponent for the 2 million molecular weight sample seems to be at the limiting value of  $-(2/3)$. Finally, the trend in the data of~\citet{Hua2006787} is clearly contradictory to that of~\citet{Noda:1968gt} with regard to the influence of solvent quality. While the former predicts a decrease in shear thinning with increasing solvent quality (most apparent for the 2 million molecular weight sample), the latter predicts an increase with increasing solvent quality (note the molecular weight in the latter case is 13.6 million).

It is challenging to make sense of the wide range of behaviour displayed by the dilute polystyrene samples  in Fig.~\ref{fig:diluteshearPS}, and the comparatively less varied behaviour observed for the dilute DNA samples considered here (displayed in Figs~\ref{fig:diluteshearDNA} and~\ref{fig:diluteshearallDNA}). It is clear that when $\eta_\text{p}/\eta_{\text{p}0}$ is plotted versus $\lambda_Z \dot \gamma$, for various values of $z$ and $c/c^*$, data collapse onto master plots, independent of temperature and concentration, is observed, (i) at all values of shear rate in the case of DNA, and (ii) only for shear rates $\lambda_Z \dot \gamma \lesssim \mathcal{O}(2)$ for polystyrene solutions. The delay in the onset of shear thinning with increasing molecular weight, which is clearly observed for DNA solutions, is not observed for polystyrene solutions, though it has been observed by~\citet{Noda:1968gt} for polystyrene in decalin and poly($\alpha$-methyl-styrene). The complex dependence of viscosity on shear rate predicted by models that include consistently-averaged, or fluctuating hydrodynamic interactions, can be invoked to understand some of the observed behaviour. According to these models, the presence of hydrodynamic interactions delays the onset of shear thinning due to finite extensibility, with the shear rate at which chains become aware of their finiteness increasing with increasing chain length. This would then explain the observed delay in the onset of shear thinning with increasing molecular weight. However, the theoretical models also predict shear thickening due to the presence of hydrodynamic interactions, but no shear thickening is apparent in the data for DNA presented here. It is worth noting that the contour lengths of most of the polystyrene samples considered here (displayed in Tables~\ref{tab:Noda} and~\ref{tab:zcPS}) are significantly smaller than the contour lengths of the DNA samples (see Table~\ref{tab:DNAprop}). The exception is the 13.6 million molecular weight sample of~\citet{Noda:1968gt}, whose contour length of 34 $\mu$m lies between that of $\lambda$-phage (16 $\mu$m) and T4 DNA (56 $\mu$m). Indeed this difference from all the other polystyrene samples might explain to some extent the reduced shear thinning of this sample, with shear thinning due to finite extensibility probably occurring only at much higher shear rates. 

The expectation that the systematic experiments carried out here for DNA solutions, and the compilation of prior experimental data for polystyrene solutions within a unified framework, would resolve many of the ambiguous observations highlighted in the introductory section has clearly not been fulfilled. At equilibrium, static and dynamic properties of dilute DNA solutions in the presence of excess salt have been shown to exhibit universal behaviour identical to that observed for synthetic polymer solutions~\cite{Pan2014339,Pan2014b}. The experiments on DNA solutions in shear flow that have been carried out here were based on a similar presupposition. However, as discussed above, significant differences in the behaviour of dilute DNA and polystyrene solutions have been observed. The question is whether these variations are due to the distinctness of DNA and polystyrene on a molecular scale. From the point of view of polymer physics, provided that neutral, linear polymer chains are sufficiently long that they have a large number of Kuhn steps, large scale properties are expected to exhibit no dependence on monomer chemistry, if the data is represented in terms of suitably normalised coordinates~\cite{DoiEd86}. A proper resolution of this question may perhaps be obtained if the conditions under which local chain details become important is better understood, and a means of doing this would be to figure out the origin of the differences in the predictions of the various polymer models, and the conditions under which these differences arise. For instance, while bead-rod models capture the second Newtonian plateau at high shear rates, which has been observed in many experiments on polystyrene solutions, bead-spring models can never predict this observation. \citet{Doyle1997251} have shown that this is because bead-spring models only account for the `Brownian' contribution to the stress tensor, while bead-rod models have a `viscous' contribution that arises because of the rigidity constraint. On the other hand, while bead-spring models predict the magnitude of the shear thinning exponent at high shear rates to be roughly 0.6 and above, which is experimentally observed for polystyrene solutions, free-draining bead-rod models predict an exponent with magnitude 0.5~\cite{Liu19895826,Doyle1997251}, which has been observed for DNA solutions~\cite{Teixeira:2005p1058,Schroeder20051967}. However, the exponent magnitude predicted by bead-rod models appears to decrease to 0.3 with the inclusion of hydrodynamic interactions~\cite{Petera19997614}. Indeed, the interplay of excluded volume interactions, hydrodynamic interactions and finite extensibility leads to the prediction of such varied and complex behaviour, that only experiments that are carefully designed to clearly delineate their separate influences, can throw light on the origin of the wide variability in experimental observations. Combined with careful computational studies, they would lead to insight on the role of local chain details, and the most appropriate polymer model for the particular conditions of the experiments.

\section{\label{sec:sdshflow} Semidilute solutions}

\subsection{Relaxation time $\pmb{\lambda_\eta}$}

\begin{figure}[tbp]  
\centering 
\resizebox{8.8cm}{!} {\includegraphics*{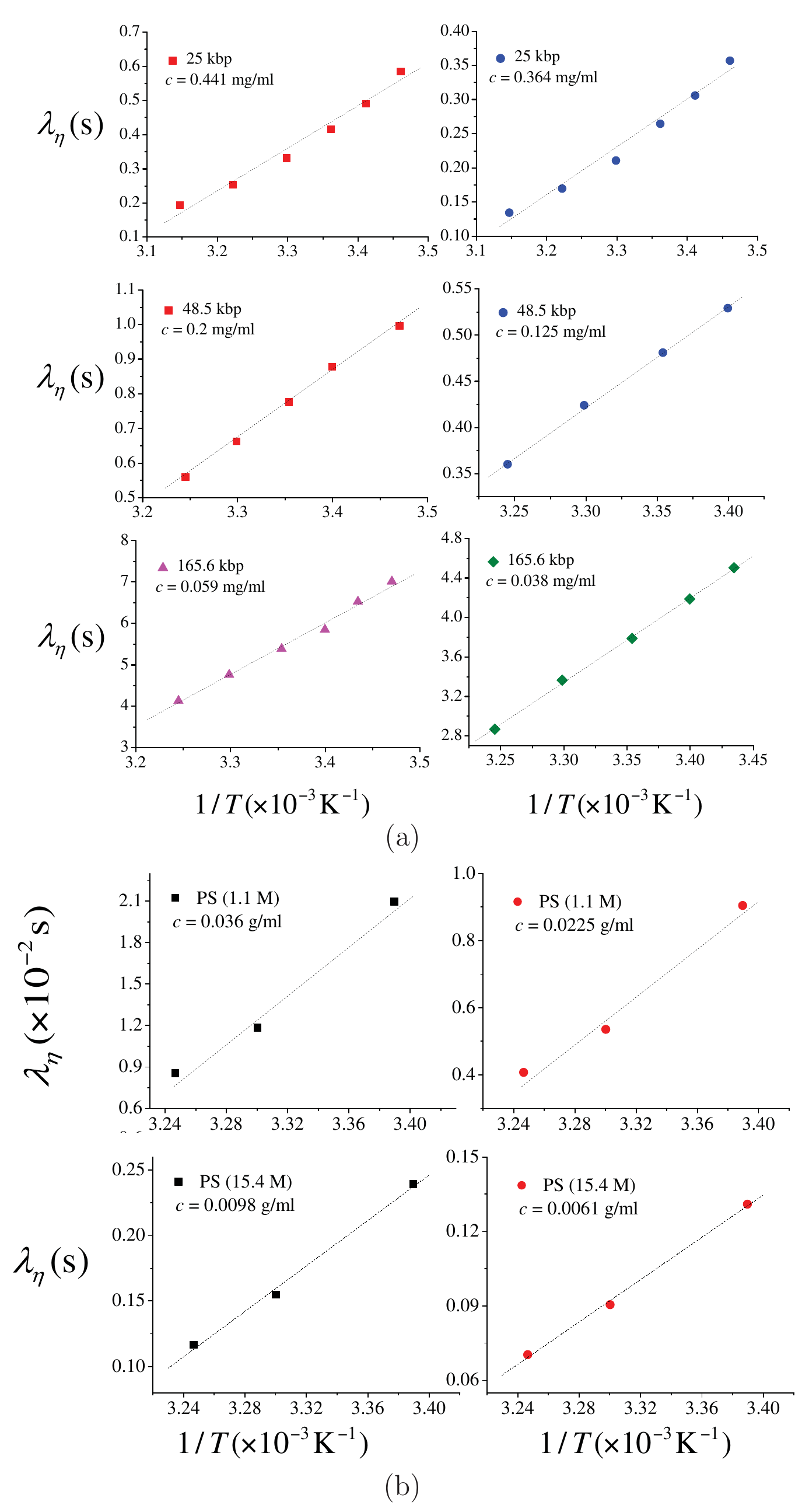}}
\vspace{-12pt}
\caption{Relaxation time, $\lambda_{\eta}$, versus inverse temperature, $(1/T)$, at fixed concentrations, $c$, for (a) 25 kbp, 48.5 kbp, and 165.6 kbp DNA, and (b) 1.1 M and 15.4 M polystyrene. The lines are  linear least squares fit to the data. Similar plots at other values of concentration are displayed in Figs.~2 and~3 of the supplementary material.}
\label{fig:LetavsinvTDNAPS}
\end{figure}

It is common, both in single molecule experiments~\cite{Hur2001421,Liu20091069,Hsiao:2017fm} and in computations~\cite{Stoltz2006137,Huang201010107,Sasmal:2017ey} involving semidilute solutions, to define the Weissenberg number in terms of the longest relaxation time $\lambda_1$. In the case of bulk rheological measurements, however, it is more convenient to use a relaxation time, $\lambda_\eta$,  defined in terms of the polymer contribution to the zero shear rate viscosity~\cite{Pan2014339},
\begin{equation}
\lambda_\eta = \frac{M \,  \eta_{\text{p}0}}{c N_A k_B T}
\label{eq:lambdaeta}
\end{equation}
Note that,  in the limit $c \to 0$, $\lambda_\eta \to \lambda_Z$. Defining the Weissenberg number in terms of either $\lambda_1$ or $\lambda_\eta$ is entirely equivalent since they both have the same dependence on concentration in the semidilute regime~\cite{Pan2014339,Hsiao:2017fm}, and their ratio is a universal constant~\cite{Pan2014339}.

The values of $\lambda_\eta$ for all the semidilute DNA and polystyrene solutions considered here can be obtained using Eq.~(\ref{eq:lambdaeta}), and the values of $\eta_0$ for these solutions listed in Tables~III and~IV of the supplementary information. As noted in Appendix~\ref{sec:HuaWu}, the ratio $([\eta]_0/M)$ is found to depend linearly on inverse temperature $(1/T)$, for dilute DNA and polystyrene solutions. Figs.~\ref{fig:LetavsinvTDNAPS}~(a) and~(b) show that, interestingly, there appears to be a linear relationship between \leta\ and $(1/T)$, for semidilute DNA and polystyrene solutions, in the range of concentrations and temperatures considered here. Additional data at other concentrations in the semidilute regime displaying similar behaviour is shown in Fig.~2 of the supplementary information.

\begin{figure}[tbp]
\begin{center}
\resizebox{8.5cm}{!} {\includegraphics*{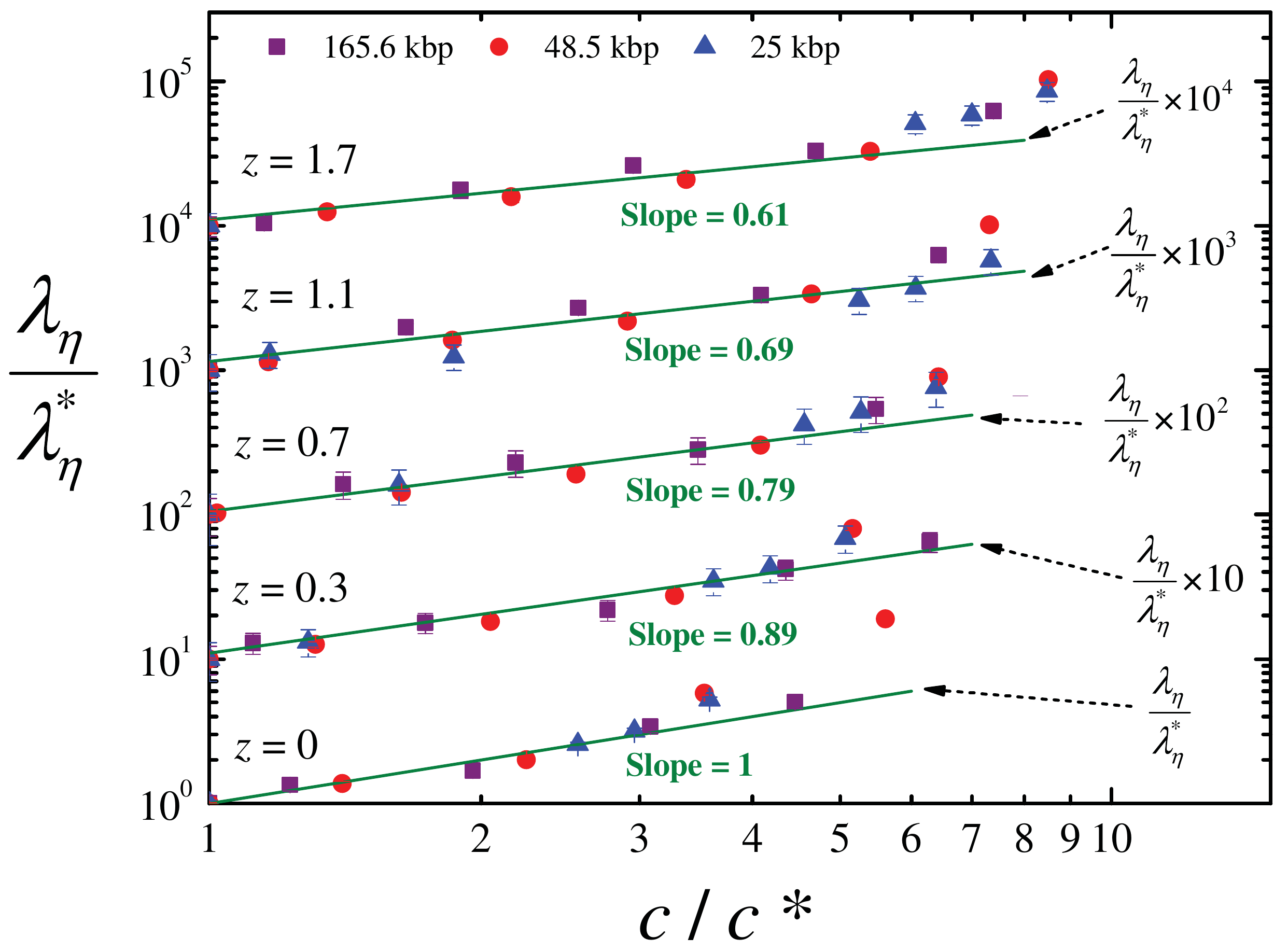}}
\end{center}
\vspace{-12pt}
\caption{Universal crossover scaling of the relaxation time $\lambda_{\eta}$. The quantity $\lambda_{\eta}^{*}$ is defined as the value of  $\lambda_{\eta}$ at $c = c^{*}$. In order to display all the measurements on a single plot, the ratios of relaxation times for the different values of $z$ have been multiplied by different fixed factors as indicated. Lines through the data have been drawn with a slope calculated using the values of \nueff\ ($z$) determined previously for the zero shear rate viscosity (see Table~IV in Ref.~\cite{Pan2014339}).}
\label{fig:letacrossover}
\vskip-15pt
\end{figure}

\citet{Pan2014339} have recently determined the universal scaling function that governs the zero shear rate viscosity of unentangled semidilute DNA solutions, in the double crossover regime defined by the variables $\{ z, c/c^*\}$. They showed that, independent of the DNA molecular weight, the zero shear rate viscosity obeys a power law, $\eta_{\text{p}0}/\eta_{\text{p}0}^* \sim (c/c^*)^{1/(3\nu_\text{eff} -1)}$, where $\eta_{\text{p}0}^*$ is the value of $\eta_{\text{p}0}$ at $c=c^*$, and the effective power law exponent, $\nu_\text{eff}$, depends on the solvent quality $z$. It follows from Eq.~(\ref{eq:lambdaeta}), that the relaxation time $\lambda_\eta$ must obey the scaling law,
\begin{equation}
\frac{\leta}{\leta^{*}} \sim \left( \dfrac{c}{c^{*}} \right)^{\tfrac{2 - 3\nueff(z)}{3\nueff(z) - 1}}
\label{eq:leta}
\end{equation}
where $\lambda_{\eta}^{*}$ is the value of  $\lambda_{\eta}$ at $c = c^{*}$. The concentration dependence of the ratio, $\leta/\leta^{*}$, for the three different molecular weights of DNA, is presented in Fig.~\ref{fig:letacrossover}, for four different values of the solvent quality $z$. Note that in order to maintain the same value of solvent quality across the various molecular weights, it is necessary to carry out experiments at the appropriate temperature for each molecular weight. Clearly, provided $z$ is the same, the data collapses onto universal power laws, independent of DNA molecular weight. The exponent $(2 - 3\nueff)/(3\nueff - 1)$ can be calculated, at each of the values of $z$, using the values of $\nueff (z)$ determined previously  by~\citet{Pan2014339}. The lines through the symbols in Fig.~\ref{fig:letacrossover} have been drawn with their slopes obtained in this manner. The dependence on $(c/c^*)$, predicted by the scaling law Eq.~(\ref{eq:leta}), is evidently obeyed.

\begin{figure}[tbp]  
\centering 
\resizebox{8.5cm}{!} {\includegraphics*{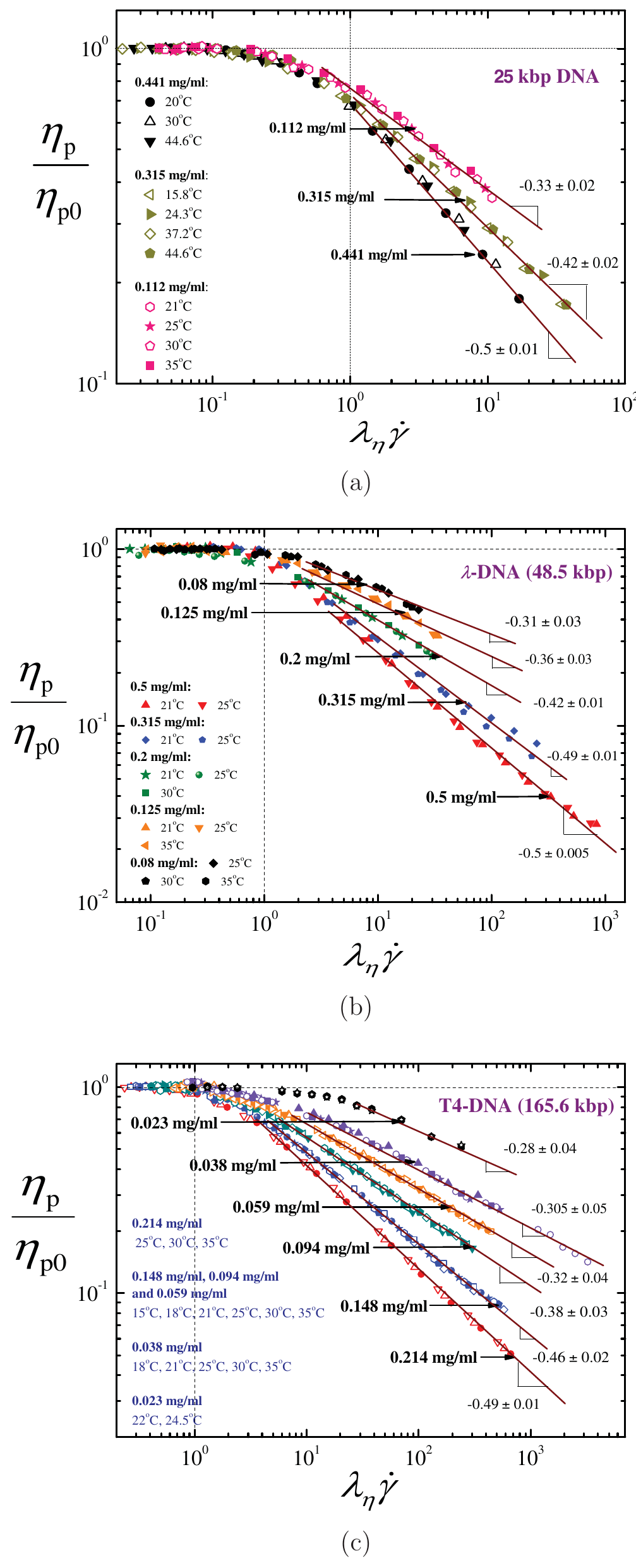}}
\vspace{-5pt}
\caption{Dependence of the scaled polymer contribution to shear viscosity, $\etap / \etapo$, on the Weissenberg number, $\leta \dot{\gamma}$, at different temperatures and concentrations, for (a) linear 25 kbp, (b) $\lambda$-phage, and (c) T4 DNA. The lines are least square fits of data in the shear thinning region.}
\label{fig:tempcollapse25lambdaT4}
\end{figure}

\begin{figure}[tbp]
\centering 
\resizebox{8.5cm}{!} {\includegraphics*{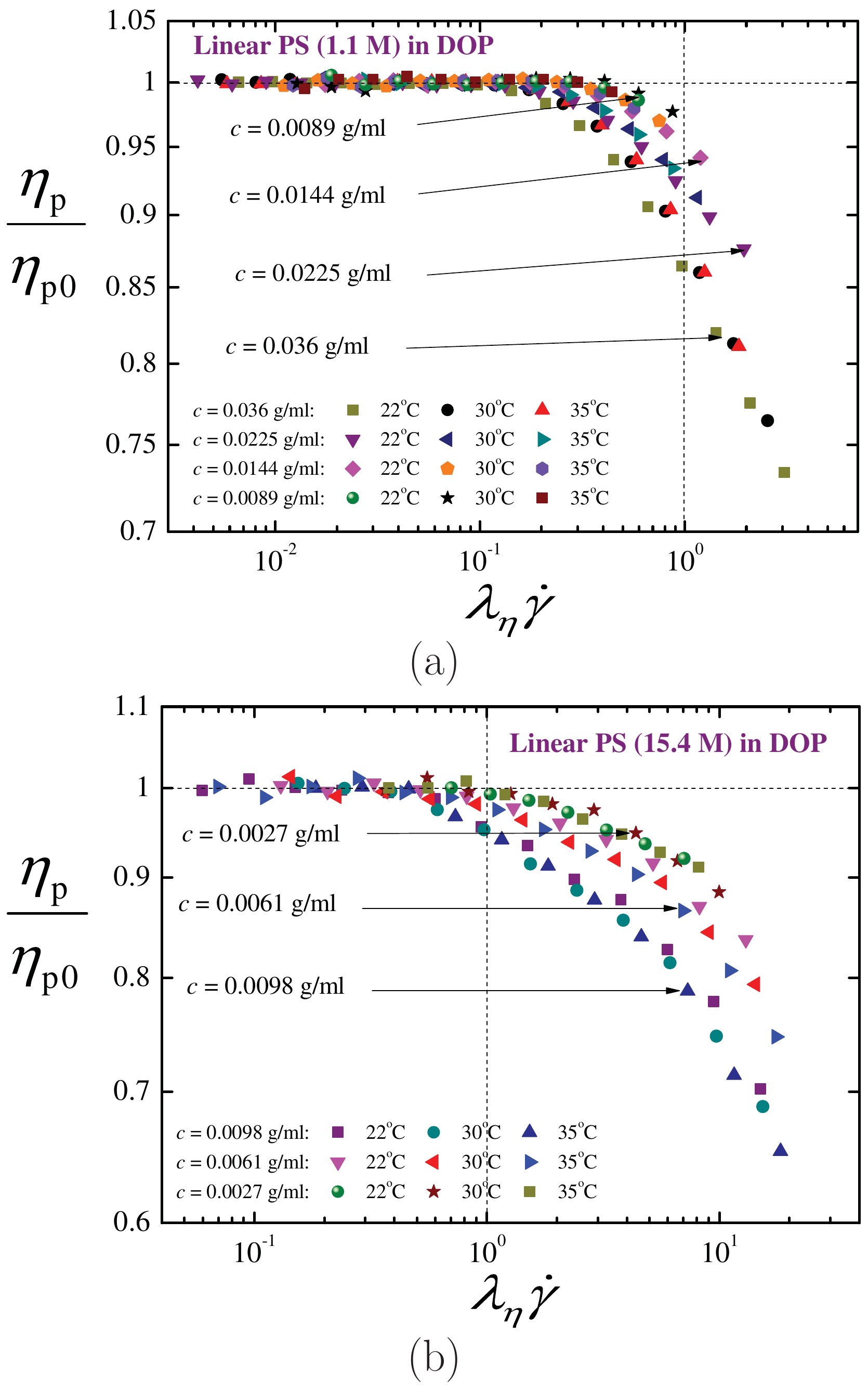}}
\caption{\label{fig:tempcollapsePS} Dependence of the scaled polymer contribution to shear viscosity, $\etap / \etapo$, on the Weissenberg number, $\leta \dot{\gamma}$, at different temperatures and concentrations, for (a)  linear 1.1 M polystyrene, and  (b) linear 15.4 M polystyrene.}
\end{figure}

\subsection{\label{sec:semidilshear} Time-temperature superposition of the dependence of viscosity on shear rate}

The dependence of the scaled viscosity $\etap / \etapo$ on the `raw' shear rate $\dot{\gamma}$, for 25 kbp, $\lambda$-phage, and T4 DNA, and the same dependence for the two polystyrene samples examined in this work, i.e., PS1.1M and PS15.4M, in the semidilute regime, are displayed in Figs.~4~(a)~to~4~(i), and Figs.~5~(a)~to~5~(h), respectively, of the supplementary material, along with a discussion of the observed behaviour. The data presented in these figures is also given in tabular form in Tables~VIII to~XIII of the supplementary material, so that they are readily available for comparison with future model predictions for the shear thinning of semidilute polymer solutions. 

In an attempt to construct master plots from the data displayed in Fig.~4 and Fig.~5 of the supplementary material, the shear rate $\dot{\gamma}$ is non-dimensionalised by multiplying it with the relaxation time $\leta$, and the consequent dependence of the scaled viscosity $\etap / \etapo$ on the Weissenberg number $\leta \dot{\gamma}$, obtained in this manner, is displayed in Figs.~\ref{fig:tempcollapse25lambdaT4}, for the 25 kbp, $\lambda$-phage and T4 DNA samples, and in Fig.~\ref{fig:tempcollapsePS}, for the 1.1 M and 15.4 M polystyrene samples, respectively. There are several striking features in these figures that we discuss in turn.

The first noticeable fact is that the various curves for $\etap / \etapo$ at
\textit{different} temperatures, but at the \textit{same} concentration collapse
on top of each other when the data is represented in this form, both for the DNA and the polystyrene solutions. This implies that using \leta\ as the relaxation time to non-dimensionalize the shear rate leads to \textit{time-temperature} superposition, i.e., all the different curves seen in
the individual subfigures of Figs.~4 and~5 of the supplementary material, collapse on to a single curve for each concentration,
independent of temperature.

\begin{figure}[tbp]
\begin{center}
\resizebox{8.5cm}{!} {\includegraphics*{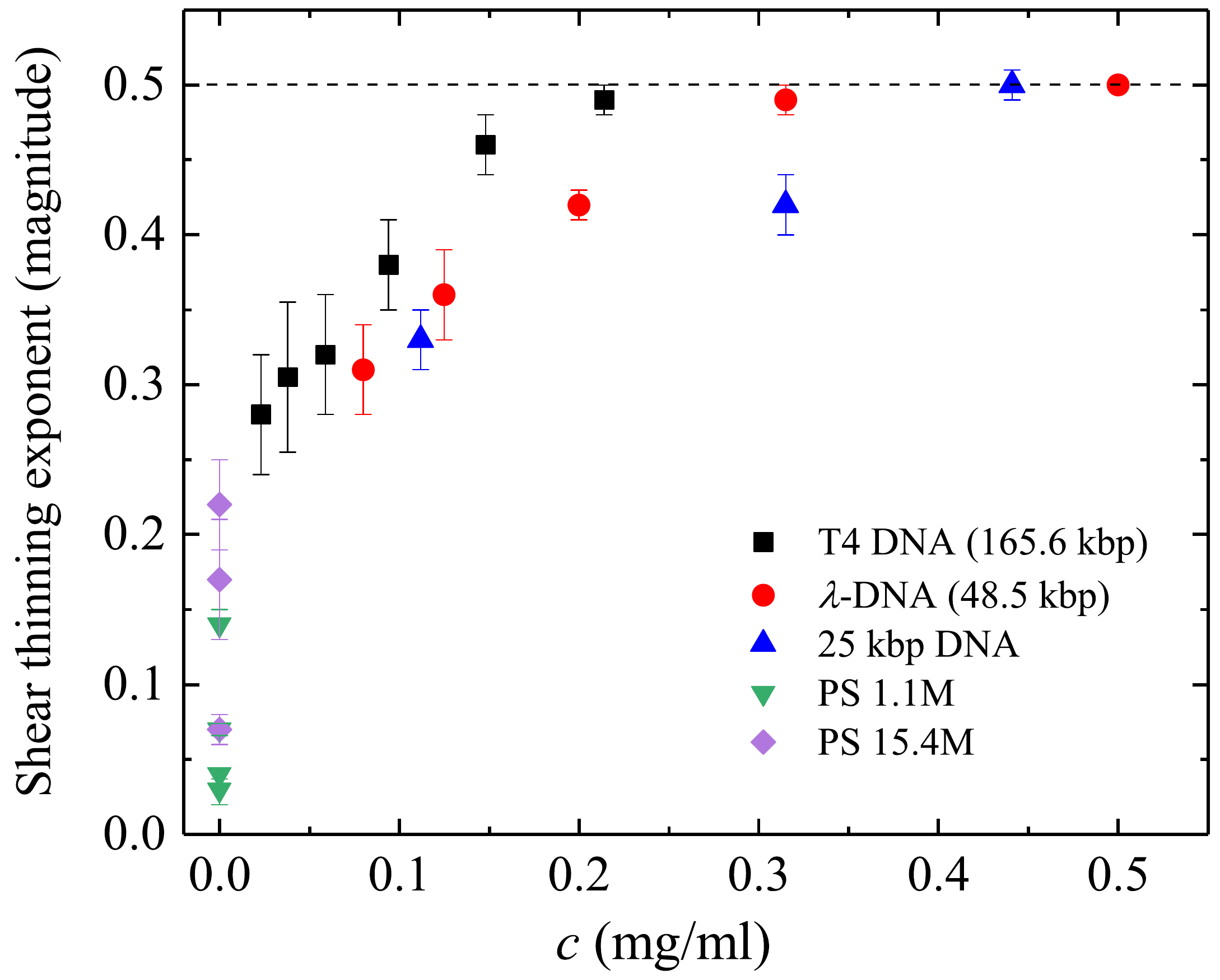}}
\end{center}
\vskip-10pt
\caption{\label{fig:slopevsconc} Slope of $\eta_{p}/\eta_{p0}$ versus $\lambda_{\eta} \dot{\gamma}$ as a function of absolute concentration $c$, obtained from the curves in Figs.~\ref{fig:tempcollapse25lambdaT4} for DNA, and from the curves in Fig.~\ref{fig:tempcollapsePS} for polystyrene.}
\end{figure}

The most significant aspect of Figs.~\ref{fig:tempcollapse25lambdaT4} for DNA and Figs.~\ref{fig:tempcollapsePS} for polystyrene, is that the data does not collapse on to master
curves, independent of \ccs\ or molecular weight, as has been observed previously in simulations~\cite{Huang201010107}. Nor is the slope in the power-law region close to $- 0.5$, as observed in the experiments of~\citet{Hur2001421} and in the simulations of~\citet{Stoltz2006137} and~\citet{Huang201010107}. Indeed, in the case of DNA, there is a significant power law regime over several decades of $\leta \dot{\gamma}$, with the magnitude of the slope increasing with increasing concentration.

The dependence of the magnitude of the slope of the $\etap / \etapo$ vs $\leta \dot{\gamma}$
curves on concentration, for DNA and polystyrene in the power law regime, 
is displayed in Fig.~\ref{fig:slopevsconc}. As can be seen, the magnitude of
the slope for the DNA samples appears to increase almost linearly before saturating to a
value of 0.5 at high concentrations. It should be noted that each of the symbols in the figure
correspond to several temperatures, and as a result, to several values of
\ccs. Consequently, while for each of the DNA samples, the asymptotic value of 0.5
is reached for increasing values of \ccs, there is no meaningful threshold
value in terms of \ccs.

The source of the lack of agreement between the current experimental observations on DNA and polystyrene, with the predictions of simulations that suggest that a concentration dependent large scale relaxation time leads to data collapse across different values of (\ccs)~\cite{Huang201010107}, is not clear to us. In the next subsection, we develop a scaling argument that suggests that a relaxation time with an alternative dependence on $(c/c^*)$ may be more appropriate to represent the universal behaviour observed in the shear flow of semidilute polymer solutions.

\begin{figure*}[tbp]
\centering
\resizebox{13cm}{!} {\includegraphics*{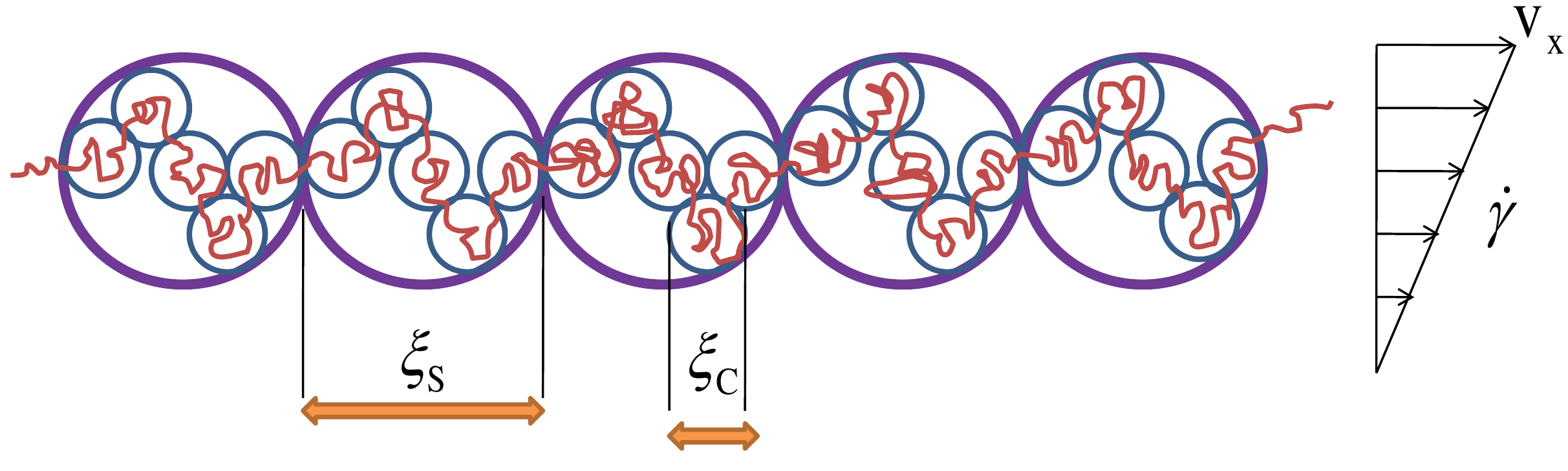}}
\caption{\label{fig:pincus}  Illustration of a linear 
polymer chain in a semidilute solution as a series of correlation 
blobs, each of size $\xi_{\mathrm{c}}$. In shear flow, the chain 
of correlation blobs breaks up into a sequence of Pincus blobs, each with a size $\xi_{\mathrm{S}}$. }
\end{figure*}

\subsection{\label{subsec:colbyrelax} The derivation of an appropriate relaxation time}

Within the blob scaling ansatz for semidilute solutions, the confirmation of a polymer chain at equilibrium consists of a series of `correlation' blobs, with size $\xi_{\mathrm{c}}$, such that chain segments below this length scale are under dilute solution conditions, while chain segments at larger length scales experience  concentrated solution conditions. As a result, within a correlation blob, chain segments interact with each other through hydrodynamic and excluded volume interactions, while both these interactions are absent between correlation blobs themselves. On length scales larger than $\xi_{\mathrm{c}}$, the absence of excluded volume interactions leads to chains obeying random walk statistics. Further, since the correlation blobs are space filling, the behaviour of an unentangled semidilute solution can be considered to be equivalent to that of a melt of Rouse chains of correlation blobs. This simple picture of semidilute solutions at equilibrium can be used to derive scaling laws for several different properties~\cite{dgen79,RubCol03}. Here, we consider semidilute solutions in the presence of shear flow, and use the blob picture to derive what the concentration dependence of a large scale relaxation time might be. Some of the equilibrium scaling laws that are useful for this purpose are reproduced in Appendix~\ref{blob} for ease of reference.

Consider a system of chains with $N$ monomers each, in the semidilute
concentration regime, undergoing simple shear flow at a shear rate
$\dot{\gamma}$. Under these conditions,
it is envisaged that the chain breaks up into a series of `Pincus'
blobs  (as depicted schematically in Fig.~\ref{fig:pincus}), where the blob size $\xi_\text{S}$ sets
the length scale at which the stretching energy of the chain segment
within the Pincus blob is equal to $\kB T$~\citep{Pincus1976386}. Clearly, with the onset of flow, it is the chain of correlation blobs that breaks up into a sequence of Pincus blobs, and one can consider
each Pincus blob to consist of $m$ correlation blobs. As a result, random walk statistics dictate that,
\begin{equation}
\label{eq:sizeblob1}
\xi_\text{S} = \xi_\text{c} m^{1/2}
\end{equation}
One can obtain an estimate of $m$, and its dependence on the shear rate
as follows. We know that the stretching energy of chain segment within
a Pincus blob is of order $\kB T$. As a result,
\begin{equation}
\label{eq:forceblob}
f \xi_\text{S} = \kB T
\end{equation}
where, $f$ is the stretching force on the chain. The stretching force
arises due to the drag exerted by the flowing fluid on the segment
within the Pincus blob. The drag force is equal to the velocity
difference across a Pincus blob times the friction coefficient of
a Pincus blob, $\zeta_\text{S}$. The velocity difference is equal
to $\dot{\gamma} \, \xi_\text{S}$, and $\zeta_\text{S} = m \, \zeta_\text{c}$ (where $\zeta_\text{c}$ is the friction coefficient of a correlation blob) since the correlation blobs obey Rouse dynamics. As a result,
from Eq.~(\ref{eq:zetac}),
\begin{equation}
\label{eq:forceblob2}
f = (\dot{\gamma} \xi_\text{S})\, m  \etas \xi_\text{c}
\end{equation}
Substituting Eq.~(\ref{eq:forceblob2}) into Eq.~(\ref{eq:forceblob}) leads to,
\begin{equation}
\label{eq:m1}
m \, \xi_\text{S}^{2} \, \etas \, \xi_\text{c} \dot{\gamma} = \kB T
\end{equation}
From Eq.~(\ref{eq:sizeblob1}), it follows that,
\begin{equation}
\label{eq:m2}
m^{2} \, \xi_\text{c}^{3} \, \etas \dot{\gamma} = \kB T
\end{equation}
or
\begin{equation}
\label{eq:m3}
m = \left( \, \left[ \frac{\kB T}{\etas \xi_\text{c}^{3}} \right] \frac{1}{\dot{\gamma}} \right)^{\tfrac{1}{2}}
\end{equation}
Using the expression for the relaxation time of a correlation blob, $\tau_\text{c}$,  in Eq.~(\ref{eq:chainrelax1}), we get,
\begin{equation}
\label{eq:m4}
m = \left( \tau_\text{c} \dot{\gamma} \right)^{-\tfrac{1}{2}}
\end{equation}
The number of correlation blobs within a Pincus blob decreases with increasing
shear rate, as might be expected, since the size of the Pincus blob decreases
with increasing shear rate.

The recognition of the inverse shear rate dependence of the Pincus blob size
is responsible for the development of a scaling model to explain shear thinning
in polymer melts by Colby and coworkers~\citep{Colby2007569}. We adopt their
arguments (with some important differences) to develop an expression for the
shear rate dependence of \etap\ in semidilute solutions, and by this means,
come up with the choice of an appropriate concentration dependent relaxation
time.

According to Rouse theory, when $\tau_\text{chain} \, \dot{\gamma} < 1$, the shear viscosity of a polymer melt is given by
the expression,
\begin{equation}
\label{eq:shvisc}
\eta = \kB T \left( \frac{\tilde{c}}{N} \right) \tau_\text{chain} \sum_{p = 1}^{N} \frac{1}{p^{2}}
\end{equation}
where $\tilde{c}$ is the number of monomers per unit volume, $N$ is the number of monomers per chain, $\tau_\text{chain}$ is the relaxation time of the chain, and the sum is carried out over the $N$ normal modes of the chain. It is appropriate to recall here that the `$p$'th mode corresponds to a segment of the chain containing ($N/p$) monomers.
Colby and coworkers~\citep{Colby2007569} argue that in a shear flow, the chain
breaks up into a series of Pincus blobs that are aligned in the flow direction,
i.e., the chain forms a `blob' pole, unlike when the chain ends
are separated by a stretching force,  in which case the chain is a directed random walk of Pincus blobs (which was the original \citet{Pincus1976386} scenario). Since the
stretching energy is less than $\kB T$ within the Pincus blob, the conformation
of the chain segment within the blob is unperturbed from its equilibrium
configuration. According to Colby and coworkers, the Rouse expression for the
viscosity (Eq.~(\ref{eq:shvisc})) only applies to chain segments within the Pincus
blob, since they are the ones contributing to viscous dissipation, while the
segments on larger length scales are stretched and store energy elastically.
\citet{Colby2007569} derive an alternative expression for the viscosity by
changing the lower bound of the summation in Eq.~(\ref{eq:shvisc}) to the mode
number corresponding to the length scale of the Pincus blob. They use
($R_\text{eq} / \xi_\text{S}$) as a measure of this mode number (where $R_\text{eq}$ is the mean size of the chain at equilibrium), and by using the shear rate dependence of $\xi_\text{S}$, they obtain a expression for the viscosity that depends on the shear rate.

In the case of semidilute solutions, for $\tau_\text{chain} \dot{\gamma} < 1$, Eq.~(\ref{eq:shvisc}) takes the form (using Eq.~(\ref{eq:ctildebyN}) for $\tilde{c} / N$),
\begin{equation}
\label{eq:shvisc2}
\eta = \kB T \left( \frac{1}{N_\text{c} \, \xi_\text{c}^{3}} \right) \tau_\text{chain} \sum_{p = 1}^{N_\text{c}} \frac{1}{p^{2}} 
\end{equation}
where $N_\text{c}$ is the number of correlation blobs in a chain. For $N_\text{c} \gg 1$, using Eq.~(\ref{eq:chainrelax3}) for $\tau_\text{chain}$, this implies,
\begin{equation*}
\eta = \frac{\kB T}{N_\text{c} \xi_\text{c}^{3}} \frac{\etas \, \xi_\text{c}^{3}}{\kB T} \, N_\text{c}^{2}
\end{equation*}
or,
\begin{equation}
\label{eq:etabyetas}
\frac{\eta}{\etas} \sim N_\text{c} \sim \left( \frac{c}{c^{*}} \right)^{\tfrac{1}{3\nu - 1}}
\end{equation}
in agreement with the expression derived by \citet{Jain2012a} for the
zero shear rate viscosity of semidilute solutions, using an alternative scaling argument.

For strong flows, as argued by Colby and coworkers~\cite{Colby2007569},
$\tau_\text{p} \dot{\gamma} < 1$ only for the modes `$p$' that lie within
a Pincus blob. Rather than ($R_\text{eq} / \xi_\text{S}$), we suggest that
it is more appropriate to use the number of Pincus blobs in a chain, $X$,
as the lower bound of the sum in Eq.~(\ref{eq:shvisc2}), since the chain
is divided into `$X$' segments by the action of the flow. For low shear rates,
$X \to 1$, since the entire chain is contained in a Pincus blob, while at
high shear rates, $X \to N_\text{c}$. Note that,
\begin{equation}
\label{eq:X}
X = \frac{N_\text{c}}{m}
\end{equation}
At high shear rates consequently, Eq.~(\ref{eq:shvisc2}) becomes,
\begin{equation}
\label{eq:shvisc3}
\eta = \kB T \left( \frac{1}{N_\text{c} \, \xi_\text{c}^{3}} \right) \tau_\text{chain} \sum_{p = X}^{N_\text{c}} \frac{1}{p^{2}} \end{equation}
Converting the sum to an integral, and carrying out the integral in
the limit $N_\text{c} \gg 1$, we get,
\begin{equation*}
\eta = \frac{\kB T}{N_\text{c} \xi_\text{c}^{3}} \, \frac{\etas \, \xi_\text{c}^{3}}{\kB T}  N_\text{c}^{2} \, \frac{1}{X} = \etas \, N_\text{c} \, \frac{m}{N_\text{c}} = \etas \, m
\end{equation*}
From Eq.~(\ref{eq:m4}), this implies that at high shear rates, for
semidilute solutions,
\begin{equation}
\label{eq:etabyetas3}
\frac{\eta}{\etas} = \left[ \tau_\text{c} \dot{\gamma} \right]^{- \tfrac{1}{2}}
\end{equation}
Using the expression for $\tau_\text{c}$ from Eq.~(\ref{eq:chainrelax2}),
we get,
\begin{equation}
\label{eq:etabyetas4}
\frac{\eta}{\etas} = \left[ \tau_{0} \, N^{3\nu} \left( \frac{c}{c^{*}} \right)^{\tfrac{3\nu}{1 - 3\nu}} \, 
\dot{\gamma} \, \right]^{-\tfrac{1}{2}}
\end{equation}
where $\tau_{0}$ is the monomer relaxation time (see Eq.~(\ref{eq:monorelax})). 

Equation~(\ref{eq:etabyetas4}) suggests that the viscosity of a semidilute polymer solution should shear thin with a power law exponent of $-(1/2)$. However, we have seen from our experimental observations that the shear thinning exponent is typically less than $- 1/2$. Nevertheless, we can conclude from Eq.~(\ref{eq:etabyetas4}) that the dependence of the relaxation time on scaled concentration
should be (\ccs)$^{3\nu/(1 - 3\nu)}$, rather than the dependence $(\ccs)^{(2 - 3\nu)/(3\nu - 1)}$
predicted for \leta. In other words, the scaling analysis here suggests that data collapse may be achieved
by using the relaxation time,
\begin{equation}
\label{eq:relaxcolby}
\lambda = \tau_{0} \, N^{3\nueff} \, \left( \frac{c}{c^{*}} \right)^{\tfrac{3\nueff}{1 - 3\nueff}}
\end{equation}
(where the effective exponent \nueff\ has been introduced to account for differences in solvent quality),  rather than \leta. This hypothesis is tested in the results presented in the next subsection. It is worth noting that, using alternative blob scaling arguments, \citet{Heo20088903} have previously obtained an identical dependence on concentration for $\lambda_\text{e}$, the Rouse time of an entanglement strand, with which they have successfully collapsed data in the concentration range, $c_\text{e} < c < c^{**}$.

Several assumptions that underlie the derivation of Eq.~(\ref{eq:relaxcolby}) would not be valid, and the scaling arguments would have to be revised, if the magnitude of the stretching force on a DNA chain in shear flow (at the typically encountered shear rates), is such that the Pincus blob size is smaller than $\xi_\text{c}$. As pointed out recently by \citet{Schroeder18} while discussing single molecule force-extension experiments, the magnitude of the force at which $\xi_\text{S} \le b_\text{k}$ is roughly 0.04 pN, which is a very small force in the context of force-extension experiments. In Appendix~\ref{pincblob}, we obtain estimates of the stretching force and the Pincus blob size in shear flow as a function of $c/c^*$, and compare the magnitude of $\xi_\text{S}$ with $\xi_\text{c}$ and $b_\text{k}$, at various values of the non-dimensional shear rate $\lambda_\eta \dot \gamma$, and demonstrate that there exists a range of shear rates in which the scaling arguments used in this section are indeed applicable.

\subsection{\label{subsec:rouse} Universal shear thinning of semidilute solutions}

\begin{figure}[tbp]
\centering
\resizebox{9cm}{!} {\includegraphics{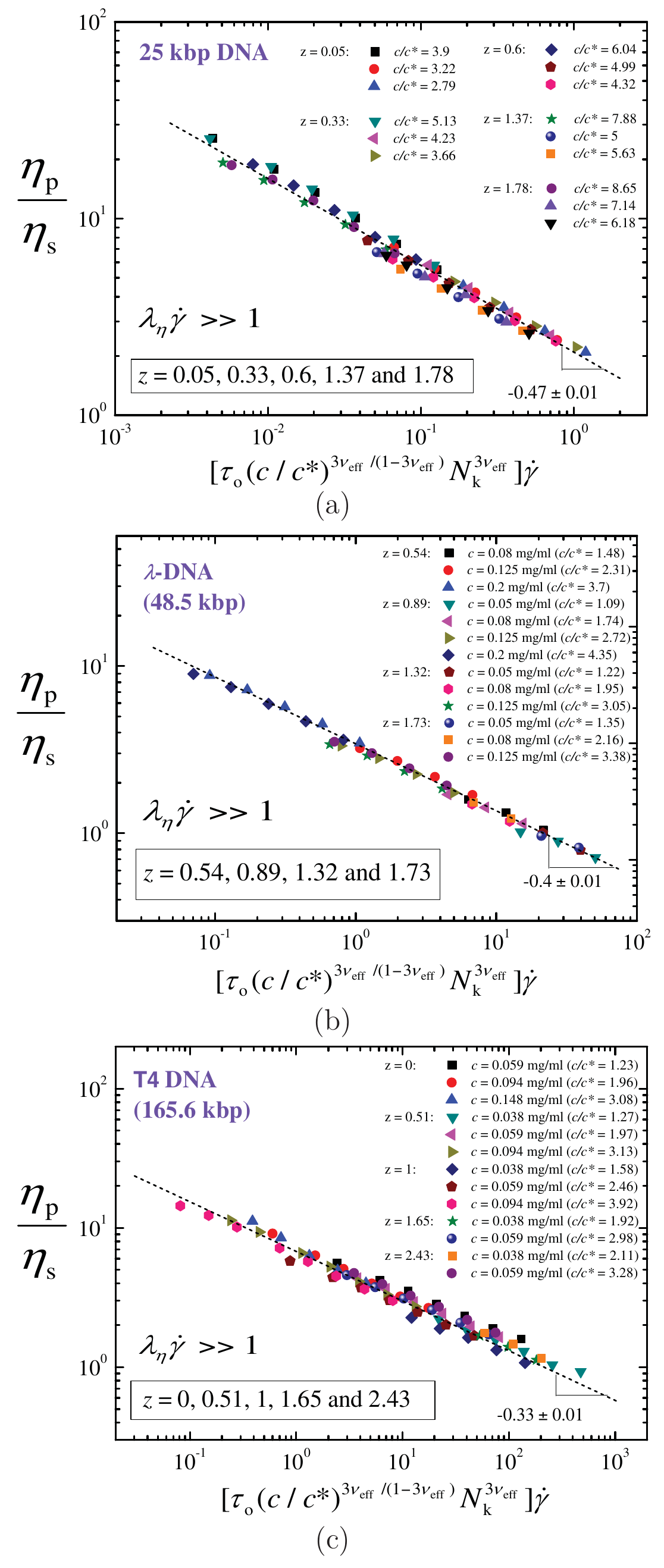}}
\caption{Universal shear thinning of semidilute DNA solutions, independent of concentration and solvent quality, at shear rates $\leta \dot{\gamma} \gg 1$, when represented in terms of a Weissenberg number that is based on the relaxation time defined by Eq.~(\ref{eq:relaxcolby}). (a) 25 kbp DNA,  (b) $\lambda$-phage DNA, and (c) T4 DNA. The value of \nueff\ corresponding to the particular value of $z$ 
has been used in the definition of the relaxation time.}
\label{fig:25kbpColby}
\end{figure}

\begin{figure}[t]
\centering
\resizebox{8.5cm}{!} {\includegraphics{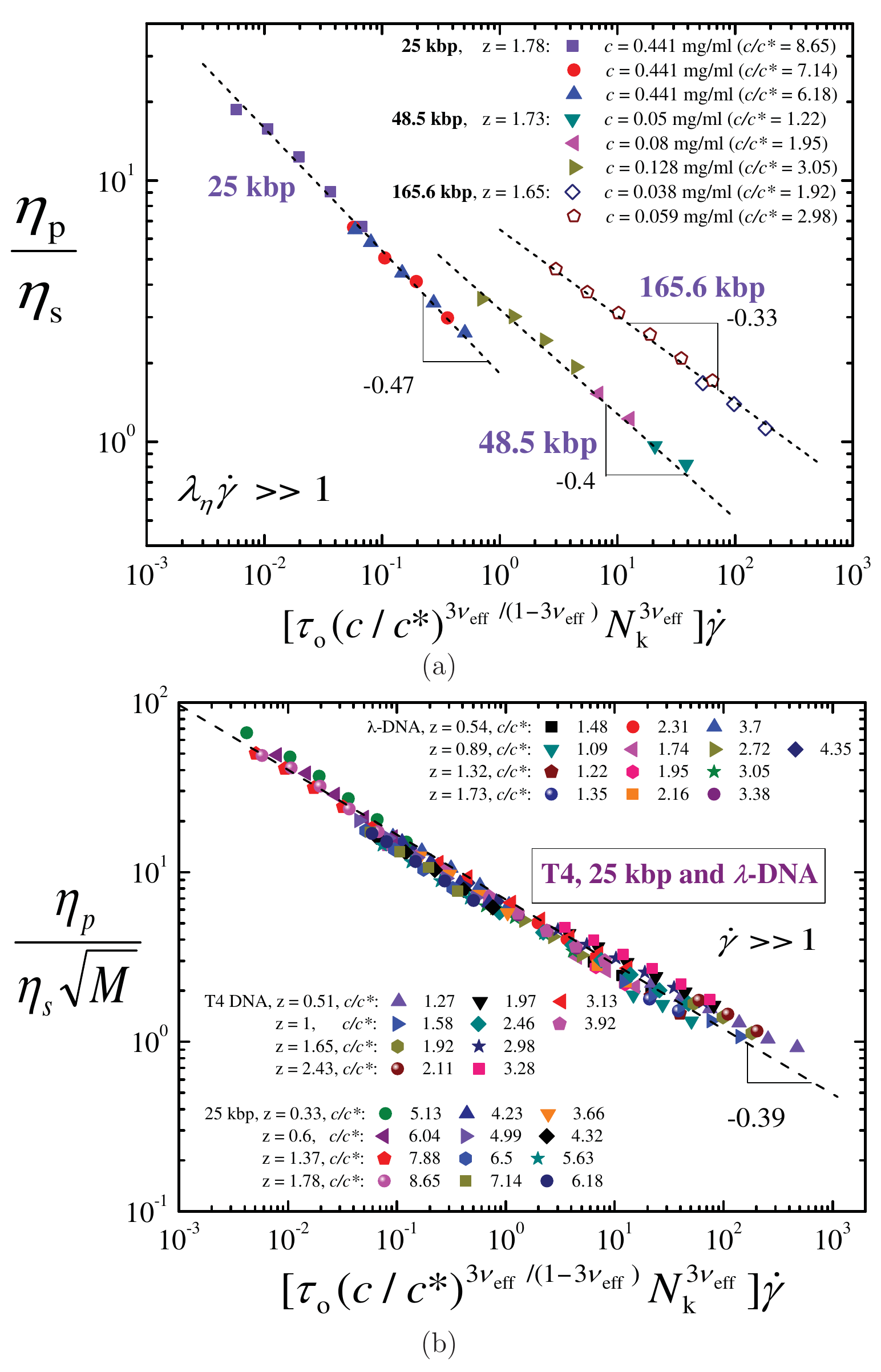}}
\caption{(a) Representative data for all three DNA molecular weights showing that the terminal slope at high shear rates decreases with increasing molecular weight. (b) Data collapse across molecular weights, concentrations and solvent quality, when the polymer contribution to viscosity \etap\ is scaled by $\etas \sqrt{M}$.}
\label{fig:Colby3MW}
\end{figure}

The `raw' viscosity versus shear rate data presented in Figs.~4 of the supplementary material, is reinterpreted in terms of the scaling variables suggested by the analysis of the previous subsection, and plotted in Figs.~\ref{fig:25kbpColby}~(a), (b) and~(c), for 25 kbp, $\lambda$-DNA and T4 DNA, respectively. It is immediately apparent that in each case, for $\leta \dot{\gamma} \gg 1$, the viscosity collapses on to master curves, independent of \ccs\ and the solvent quality $z$. It should be noted that the value of \nueff\ corresponding to the particular value of $z$ has been used in these plots. The slope of the shear thinning region seems to vary with molecular weight, with the magnitude decreasing as the molecular weight increases. This can be seen more clearly in Fig.~\ref{fig:Colby3MW}~(a), where representative data for all the three DNA molecular weights are plotted side by side. It is possible that the experimental data lies in a
crossover region, and that the Weissenberg numbers at which asymptotic shear
thinning with exponent -0.5 is observed, have not been explored for $\lambda$-phage and T4 DNA. 

We find that dividing the polymer contribution to viscosity \etap\ by $\etas \sqrt{M}$ leads to a collapse of data across molecular weights, concentrations and solvent quality, as shown in Fig.~\ref{fig:Colby3MW}~(b). This result is not obvious from the scaling arguments developed in section~\ref{subsec:colbyrelax} above. Nevertheless, it is striking, and awaits a satisfactory explanation.

An attempt to use the relaxation time given by Eq.~(\ref{eq:relaxcolby}) to collapse
the data for the two polystyrene solutions (presented in Figs.~5 of the supplementary material) is shown in Figs.~\ref{fig:ColbyPSattheta}~(a) and (b).
Clearly, all the data is in the regime prior to the onset of significant shear thinning,
and no conclusions can consequentially be drawn from the lack of data collapse,
which is expected only in the power law regime. More data at higher shear rates
is necessary to resolve this issue.

\section{\label{sec:conc} Conclusions}

\begin{figure}[t]
\centering
\resizebox{8.5cm}{!} {\includegraphics{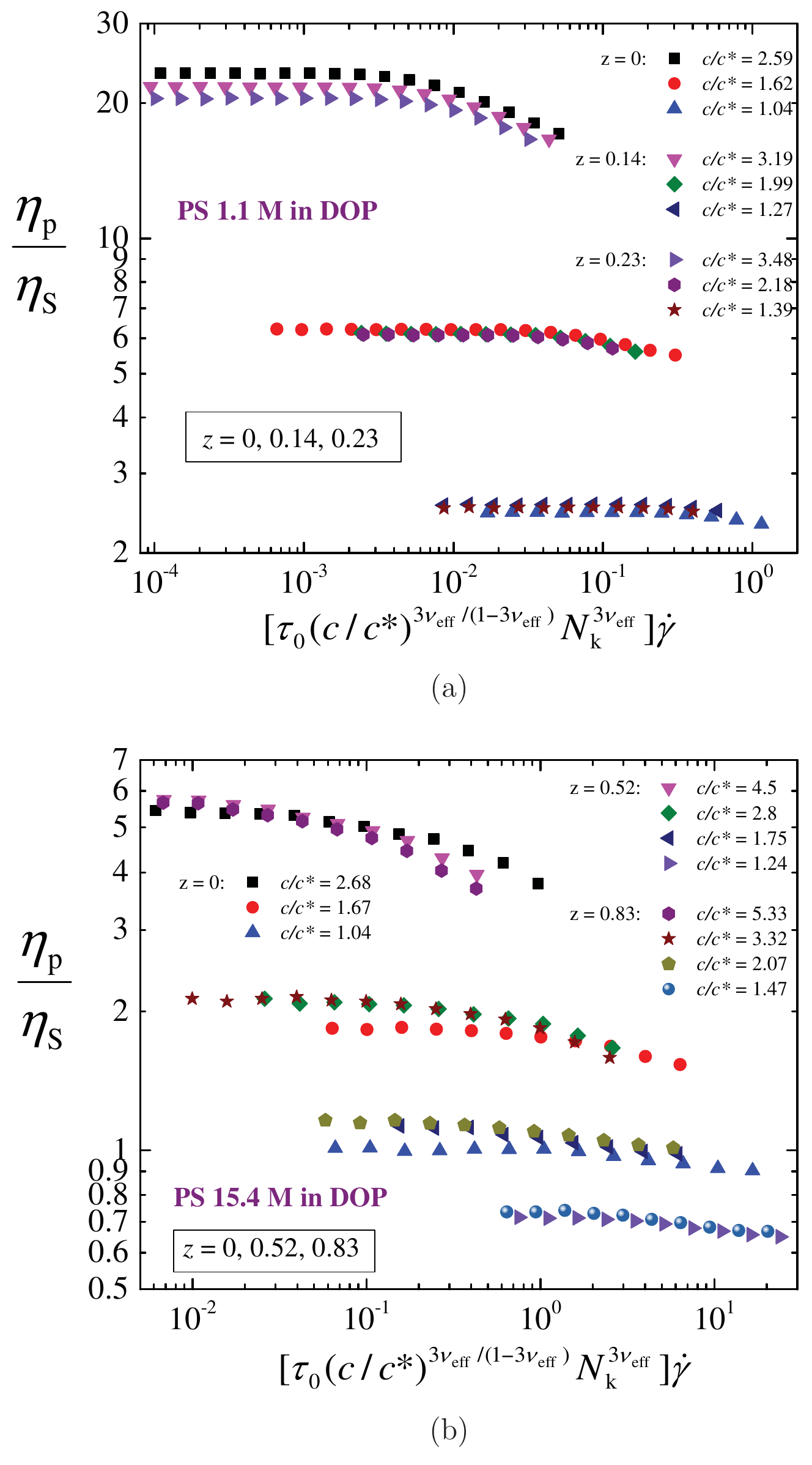}}
\caption{Shear rate dependence of the viscosity of semidilute solutions of linear polystyrene in DOP, for two different molecular weights: (a) 1.1 M and (b) 15.4 M, when represented in terms of a Weissenberg number that is based on the relaxation time defined by Eq.~(\ref{eq:relaxcolby}). A value of $\nueff = 0.5$ has been used at $T_{\theta} = 22\degC$ for $\theta$-solvents (corresponding to $z$ = 0).}
\label{fig:ColbyPSattheta} 
\end{figure}

Measurements of the viscosity of dilute and semidilute unentangled DNA solutions in simple shear flow have been carried out over a wide range of temperatures and concentrations. While similar measurements have been made for semidilute unentangled polystyrene solutions, previously reported data on dilute polystyrene solutions has been adapted for the central purpose of this work, which is to examine (in all these cases), if universal behaviour is exhibited when a proper choice of scaling variables is made. In particular, since it is well established that the dependence of equilibrium static and dynamic properties of polymer solutions on the variables $\{T, c, M\}$, becomes universal when represented in terms of the solvent quality $z$, and the scaled concentration $c/c^*$, we examine if extending this space by adding the Weissenberg number $\lambda \dot \gamma$ as a non-dimensional variable, leads to a universal representation of non-equilibrium behaviour as well. 

In order to pursue this goal, we have firstly determined the values of $z$ and $c/c^*$ for all the solutions considered here. While this has been done previously by us for the three DNA molecular weights considered here~\cite{Pan2014339,Pan2014b}, their estimation for polystyrene solutions has been discussed here in detail (see Appendix~\ref{sec:HuaWu}, and Tables~\ref{tab:zc}, \ref{tab:Noda}, \ref{tab:zcPS}, and~\ref{tab:PSzc}). 

For dilute solutions, the Weissenberg number has been defined in terms the large scale relaxation time $\lambda_Z$, which is based on the zero shear rate intrinsic viscosity (see Eq.~(\ref{eq:lambdaZ})). Using previously reported data on intrinsic viscosities of DNA and polystyrene solutions, and estimating them with a systematic procedure when they are not available (as detailed in Appendix~\ref{sec:HuaWu}), we have calculated $\lambda_Z$ for all the solutions considered here (see Tables~\ref{tab:ivisc} and~\ref{tab:iviscPS}). By determining the zero shear rate viscosity, $\eta_{p0}$, from the shear rate independent viscosity data at low shear rates (see Tables~I and~II of the supplementary material), it becomes possible to plot the ratio $\eta_{p}/\eta_{p0}$ as a function of the Weissenberg number $\lambda_Z \dot \gamma$, for both the DNA and the polystyrene solutions, at various values of $z$ and $c/c^*$, in the dilute regime (see Figs.~\ref{fig:diluteshearDNA} and~\ref{fig:diluteshearPS}).  

When represented in this form,  the data for dilute DNA solutions clearly indicates the existence of master plots independent of $z$ and $c/c^*$. However, the onset of shear thinning occurs at higher values of $\lambda_Z \dot \gamma$ with increasing molecular weight of the DNA chains. Interestingly, the shear thinning exponent in all cases is less than $-(2/3)$, which is the known limiting value for finitely extensible chains. The behaviour is more complex in the case of dilute polystyrene solutions. While data from different groups collapse on top of each other for $\lambda_Z \dot \gamma \lesssim \mathcal{O}(2)$, there is considerable variation in the results reported at higher shear rates, both in terms of the dependence on solvent quality and on molecular weight. It is hoped that future work will throw light on the unresolved issues that have been clearly identified here through the adoption of a common framework for describing equilibrium and non-equilibrium behaviour. 

In spite of the wealth of analytical and simulation studies of dilute polymer solutions that have been carried out so far, there is currently no polymer model that is able to consistently predict all aspects of the complex rheological behaviour that is observed. There is clearly a need for more systematic studies in determining the influence of excluded volume and hydrodynamic interactions on the predictions of bead-rod models, along with a need for a systematic comparison of bead-spring and bead-rod models. Such studies would help to elucidate the conditions under which it is meaningful to coarse-grain chain segments to springs, and identify the most appropriate model that should be used for a given set of experimental conditions.

For semidilute solutions, a large scale relaxation time, $\lambda_\eta$, based on the zero shear rate viscosity, has been defined (see Eq.~(\ref{eq:lambdaeta})), that has the same dependence on the scaled concentration $c/c^*$, as the longest relaxation time, $\lambda_1$, used previously in simulation studies~\cite{Stoltz2006137,Huang201010107,Sasmal:2017ey}. Values of  $\eta_{0}$ for semidilute unentangled DNA and polystyrene solutions have been determined from the shear rate independent viscosity data at low shear rates (see Tables~III and~IV of the supplementary material). In the double crossover regime defined by the variables $\{ z, c/c^*\}$, the expectation that the ratio $\leta/\leta^{*}$ has a power law dependence on $c/c^*$, with an effective exponent $\nueff$ that depends on $z$ (see Eq.~(\ref{eq:leta})), is shown to be satisfied for semidilute unentangled DNA solutions (see Fig.~\ref{fig:letacrossover}).

The use of a Weissenberg number based on $\lambda_\eta$ for semidilute unentangled DNA and polystyrene solutions, collapses data at different temperatures but at the same concentration,  on top of each other, establishing the existence of time-temperature superposition (see Figs.~\ref{fig:tempcollapse25lambdaT4} and Fig.~\ref{fig:tempcollapsePS}). The failure to collapse data for different concentrations is contrary to predictions by recent simulations~\cite{Huang201010107}, which find such a collapse for a Weissenberg number based on $\lambda_1$. 

A scaling theory for semidilute unentangled solutions that pictures a polymer chain as a blob pole of Pincus blobs in the presence of flow, leads to the definition of a Weissenberg number which is based on a relaxation time whose concentration dependence arises from the dependence on $c/c^*$ of a single correlation blob. Remarkable data collapse, independent of solvent quality and concentration, is obtained  at high shear rates when the shear rate dependence of viscosity is interpreted in terms of such a Weissenberg number. 

There is clearly a need for theoretical analysis and detailed mesoscopic simulations that go beyond the simple scaling theory presented here in order to understand the complex and rich behaviour displayed by semidilute unentangled polymer solutions. The experimental data reported here provide benchmark results for the validation and elucidation of future theoretical and computational studies.

\section*{Acknowledgements}

This research was supported under Australian Research Council's Discovery
Projects funding scheme (project number DP120101322). We are grateful to
Douglas E. Smith and his group in the University of California, San Diego,
and Brad Olsen, MIT, for originally synthesizing and sending us the agar
stab culture of \emph{E. coli} containing the 25 kbp DNA fragment. The
authors would like to thank M. K. Danquah (formerly at Monash University)
for providing laboratory space for storing DNA samples, and for the
instruments and facilities for extracting DNA. We also acknowledge the
funding and general support from IITB-Monash Research Academy.

\appendix


\section{\label{sec:HuaWu} Determination of $\pmb{\{ z, c^*, \lambda_Z \} }$ for the dilute polystyrene solutions used in~\citet{Hua2006787}}

\citet{Hua2006787} have reported viscosity measurements for four different polystyrene molecular weight samples dissolved in dioctyl phthalate (DOP). The values of the molecular weights, and the sample concentrations are reproduced in Table~\ref{tab:PSprop}, with the nomenclature used in their work. The DOP solvent is reported to have a viscosity of 53 mPa.s at room temperature, and is considered a $\theta$-solvent for polystyrene at 22\degC~\citep{Brandrup1999}. Viscosity measurements have been carried out with a Couette rheometer at the $\theta$-temperature, and three temperatures in the good solvent regime. The detailed procedure to represent the measurements of~\citet{Hua2006787} in terms of the non-dimensional variables $(z, c/c^*, \lambda_Z \dot \gamma)$ is discussed below. 

\begin{table}[tbh]
  \caption{\label{tab:PSprop} Molecular weights and concentrations of the polystyrene samples used in the experiments of~\citet{Hua2006787}. }
\vskip5pt
\begin{small}
\setlength{\tabcolsep}{5pt}
{\def\arraystretch{1.1}
\begin{tabular}{ c  c  c  c}
\hline
\multirow{2}{*}{Sample}     & Molecular weight 
            & \multicolumn{2}{c}{Concentration} \\  
     &  $M$ ($\times 10^{6}$ g/mol)
            &  $c_w$ (wt\%) &  $c$ (g/ml) \\
\hline
\hline
PS1
            & 0.55
            & 0.40
            & 0.0046
\\
PS2
            & 0.68
            & 0.45
            & 0.0048
\\
PS3
            & 0.93
            & 0.50
            & 0.0054
\\
PS4
            & 2.0
            & 0.30
            & 0.0028
\\
\hline
\end{tabular}
}
\end{small}
\end{table}

\begin{figure}[tb]  
\centering 
\resizebox{8.5cm}{!} {\includegraphics{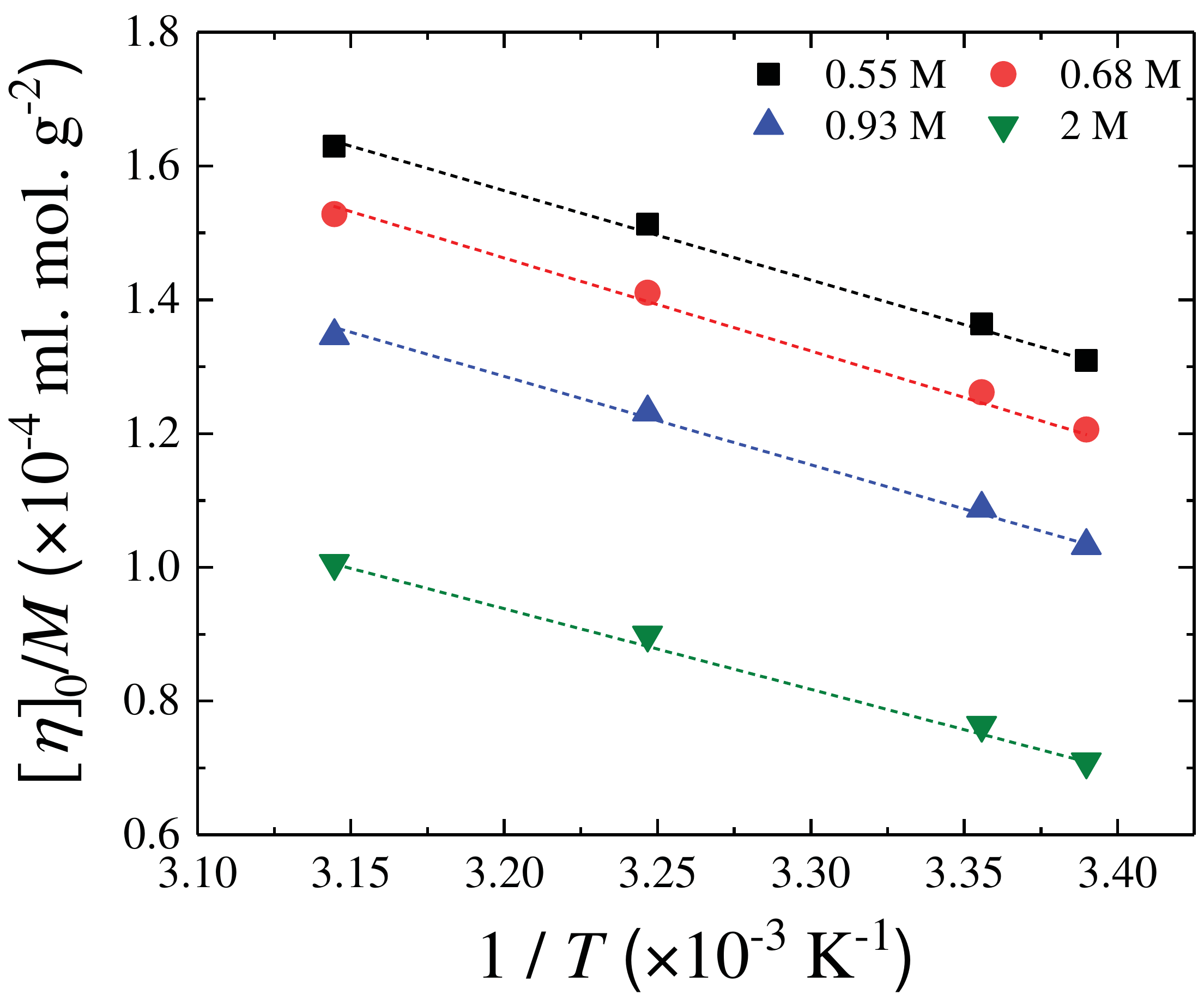}}
\vspace{-12pt}
\caption{Temperature dependence of $\ivisc_0/M$ for the four polystyrene solutions PS1 to PS4 used by~\citet{Hua2006787}. The lines are the linear least squares fit to the data.}
\label{fig:PS_intvis_vs_invT}
\end{figure}

It is necessary to know the value of the chemistry dependent constant $k$ in Eq.~(\ref{eq:z}) in order to calculate the solvent quality $z$ for the polymer solutions used by~\citet{Hua2006787}. An estimate of $k$ can be obtained from the intrinsic viscosity data reported by~\citet{Hua2006787} as follows. The dependence of intrinsic viscosity on the molecular weight of the four polystyrene samples PS1 to PS4 is plotted in Fig.~1 of Ref.~\cite{Hua2006787} at two temperatures, $T=\Ttheta = 22\degC$ and $T = 35\degC$. Data extracted from this figure for  $\Ttheta = 22\degC$ (corresponding to $z=0$) is reproduced in the first row of Table~\ref{tab:iviscps}. As reported by~\citet{Hua2006787}, the values obey the molecular weight scaling, $\ivisc_{0,\theta} \sim M^{0.52}$. As discussed in section~\ref{ultra}, values of $\ivisc_0$ at any other temperature can be obtained from the expression, $\ivisc_0 = [\eta]_{0,\theta} \, [\alpha_{\eta}(z) ]^{3}$. In particular, one can determine the intrinsic viscosity at $T = 35\degC$ for each of the four different molecular weights by guessing a value of the constant $k$, calculating the value of $z$ from Eq.~(\ref{eq:z}), and using the universal equation for $\alpha_{\eta}(z)$ obtained previously by BD simulations~\citep{Pan2014b}. The values calculated by this procedure can then be compared with the values reported in Ref.~\cite{Hua2006787} for $T = 35\degC$ in order to refine the guess for the constant $k$. Using this trial and error procedure, we find that the value $k= 0.0055 \, \text{(g/mol)}^{-1/2}$ leads to values of intrinsic viscosity that are within a few percent of values extracted from Fig.~1 in Ref.~\cite{Hua2006787}. As a result, we take this to be the value of $k$ for the polystyrene solutions used in Ref.~\cite{Hua2006787}. Once the value of $k$ is known, the solvent quality and intrinsic viscosity at any other temperature and molecular weight can be calculated as described above. The values of $z$ and $\ivisc_0$ calculated using this procedure are given in Table~\ref{tab:iviscps}, with an asterisk indicating that at those temperatures, $\ivisc_0$ has been plotted versus $M$  in Fig.~1 of Ref.~\cite{Hua2006787}. Figure~\ref{fig:PS_intvis_vs_invT} displays a plot of $\ivisc_0/M$ for the four polystyrene solutions as a function of $1/T$. The linear dependence is in line with the recent work of~\citet{Rushing20032831} who have shown that this ratio scales linearly with $1/T$ for a number of different polymer-solvent systems, with a slope that is independent of molecular weight. A similar linear dependence was shown in our earlier work to exist for the 25 kbp and T4 DNA solutions~\cite{Pan2014b}.

\begin{table*}[tbh]
\addtolength{\tabcolsep}{0pt}
\caption{\label{tab:iviscps} Solvent quality $z$, intrinsic viscosity $\ivisc_0$ (in ml/mg),  overlap concentration $c^*$ (in g/ml), scaled concentration $c/c^*$, and relaxation time $\lambda_Z$ (in s) for dilute polystyrene samples PS1 to PS4 used in Ref.~\cite{Hua2006787}, at various temperatures $T$. An asterisk indicates that $\ivisc_0$ has been plotted versus $M$ at those temperature in Fig.~1 of Ref.~\cite{Hua2006787}, while the values of $\ivisc_0$ at remaining temperatures are estimated here.}
\vspace{2pt}
\centering
\begin{small}
\setlength{\tabcolsep}{2.5pt}
{\def\arraystretch{1.3}
\resizebox{\textwidth}{!}
{\begin{tabular}{l c c c c c c c c c c c c c c c c c c c c c c c }
\hline
$T$ & \multicolumn{5}{c}{PS1 (0.55M)} &  & \multicolumn{5}{c}{PS2 (0.68M)} &  & \multicolumn{5}{c}{PS3 (0.93M)}  &  & \multicolumn{5}{c}{PS4 (2M)}\\
\cline{2-6} \cline{8-12}  \cline{14-18} \cline{20-24}
 (\degC) & $z$ & $[\eta]_0$ &   $c^*$ &   $c/c^*$ & $\lambda_Z$ & & $z$ & $[\eta]_0$ &   $c^*$&   $c/c^*$ & $\lambda_Z$& &$z$& $[\eta]_0$ &   $c^*$ &   $c/c^*$ & $\lambda_Z$& & $z$ & $[\eta]_0$ &   $c^*$ &  $c/c^*$ & $\lambda_Z$ \\ 
\hline
\hline
$22^*$  & 0  & 0.072 & 0.021	& 0.223 &	0.0011   & & 0 & 0.082 & 0.018	& 0.266 & 0.0015 & & 0  & 0.096  & 0.016	& 0.350 & 0.0024 & & 0  & 0.142 & 0.011	& 0.269 & 0.0076  \\

25 & 0.041  & 0.075 & 0.019	& 0.239 & 0.0009 &  & 0.046 & 0.086 & 0.017 & 0.287 & 0.0013 & &  0.053  & 0.101 & 0.014 & 0.383 & 0.0021 & &  0.078  &  0.153 & 0.009 & 0.305 & 0.0067 \\

$35^*$ & 0.17  & 0.083 & 0.0163	& 0.283 & 0.0006 &  & 0.19 & 0.096 & 0.014 & 0.344 & 0.0008 & &   0.22  & 0.115 & 0.012 & 0.469 & 0.0013 & &  0.33  & 0.180 & 0.007 & 0.395 & 0.0045 \\

45 & 0.30 & 0.090 & 0.014	& 0.318 & 0.0004  & & 0.33 & 0.104 & 0.012	& 0.390 & 0.0005 & &   0.38 & 0.125 & 0.010	& 0.536 & 0.0009 & &  0.56  & 0.201 & 0.006	& 0.464 & 0.0031 \\
\hline
\end{tabular}}
}
\end{small}
\end{table*}

The concentration of the dilute polystyrene solutions used by~\citet{Hua2006787} has been reported in terms of weight percent (reproduced here as $c_w$ in Table~\ref{tab:PSprop}), rather than in g/ml, which are the units used in the current work, represented by $c$. In order to convert to the latter, one requires the density of the solution $\rho_\text{soln}$, since $c = \rho_\text{soln} \, c_w$. The value of $\rho_\text{soln}$ has not been reported in Ref.~\cite{Hua2006787} for the various solutions. However, in Table~2 of their paper, \citet{Hua2006787} have reported their estimate of the overlap concentration $c^*_w$, at $\Ttheta = 22\degC$,  and at one temperature in the good solvent regime, $T = 35\degC$, which they state has been calculated using the expression, $c^*_w = 1/ (\rho_\text{soln} \, \ivisc_0)$. Since the values of $\ivisc_0$ are known at these temperatures (see Table~\ref{tab:iviscps}), one can find the values of $\rho_\text{soln}$ that lead to the reported values of $c^*_w$. We have used this procedure to find estimates of $\rho_\text{soln}$ for all four solutions PS1 to PS4, which lead to values of $c^*_w$ within a few percent of the reported values. These values of $\rho_\text{soln}$ have been used to calculate the values of concentration $c$ in g/ml, listed in Table~\ref{tab:PSprop}. 

It is possible to use the intrinsic viscosity data for the solutions PS1 to PS4 to get more accurate estimates of the overlap concentration $c^*$ than those calculated by~\citet{Hua2006787}, by making use of the universal viscosity ratio $U_{\eta R}$ defined by the expression~\cite{Pan2014b}, 
\begin{equation}
\label{eq:uetar}
U_{\eta R} \equiv \frac{ \ivisc_0 M}{ (4\pi/3) \Rg^{3} \Na}  = \frac{6^{\frac{3}{2}}}{(4\pi/3)}\, \frac{\Phi}{\Na} 
\end{equation}
where $\Phi$ is the Flory-Fox constant~\cite{RubCol03}, $\Phi =  \ivisc_0 M /6^{\frac{3}{2}} \Rg^{3}$. It is straight forward to show from Eq.~(\ref{eq:uetar}) and the definition of $c^*$ that,
\begin{equation}
\label{eq:cstar}
c^* (z) = \frac{U_{\eta R} (z) }{\ivisc_0 (z)} 
\end{equation}
\citet{Pan2014b} have shown recently that the dependence of the universal ratio $U_{\eta R}$ on the solvent quality $z$, obtained from BD simulations and experimental measurements on synthetic polymer solutions and on 25 kbp and T4 DNA solutions, can be described very well by the expression,
\begin{equation}
\label{eq:uer1}
{\Uer} =  {U_{\eta R}^{\theta}} \left(  \frac{1 + a_{\eta} z + b_{\eta} z^{2} + c_{\eta} z^{3}}{1 + a_{g} z + b_{g} z^{2} + c_{g} z^{3}}\right)^{3m/2}
\end{equation}
where $U_{\eta R}^{\theta} = 1.49$, and the suffixes on the parameters $a$, $b$ and $c$ indicate constants in the relevant expressions for the universal swelling functions $\alpha_\text{g}(z)$ and $\alpha_\eta(z)$, given in sections~\ref{subsec:DNAzc*} and~\ref{ultra}, respectively. Since we know the solvent quality and the intrinsic viscosity for the four polystyrene solutions at the various temperatures, the overlap concentration $c^*$ can be calculated from Eqs.~\ref{eq:cstar} and~\ref{eq:uer1}. The scaled concentration $c/c^*$ can then be found using the values of $c$ listed in Table~\ref{tab:PSprop} for each of the solutions. Values of $c^*$ and $c/c^*$ calculated in this manner are listed in Table~\ref{tab:iviscps}. 

The large scale dilute solution relaxation time $\lambda_Z$ (defined in Eq.~(\ref{eq:lambdaZ})) for the polystyrene solutions PS1 to PS4 can be obtained once the solvent viscosity $\eta_s$ and its dependence on temperature is known. The latter has not been reported by~\citet{Hua2006787}, who give the viscosity at room temperature to be 53 mPa.s. The dependence on temperature of the viscosity of dioctyl phthalate solutions has been shown by~\citet{Pilar2003} to obey the Vogel-Fulcher equation, $\ln \eta_s = A+B/(T+C)$, with the constants $A=-2.4617$, $B=707.86$, $C=84.543$, and temperature $T$ in degrees centigrade. This expression leads to $\eta_s=53$ mPa.s at $T=25.5 \degC$. Calculating values of $\eta_s$ at all temperatures of interest here with this expression, and using the appropriate values of $\ivisc_0$, the relaxation times $\lambda_Z$ for the dilute polystyrene solutions used in Ref.~\cite{Hua2006787}, are listed in Table~\ref{tab:iviscps}.

\section{ \label{blob} Blob scaling laws for semidilute solutions at equilibrium}

The postulate that a polymer chain in a semidilute solution at equilibrium is a sequence of correlation blobs can be used to derive some simple scaling expressions for various solution properties~\cite{dgen79,RubCol03}. Some of these expression are reproduced here since they are useful for the derivations in section~\ref{subsec:colbyrelax}.

If a chain consists of $N_\text{c}$ correlation blobs that obey random walk statistics, then
\begin{equation}
R_\text{eq} \sim \xi_\text{c} N_\text{c}^{\frac{1}{2}}
\end{equation}
where $R_\text{eq}$ is the mean size of the chain at equilibrium. If $\tilde{c}$
is the number of monomers per unit volume, and $N$ is the number of monomers per chain, then ($\tilde{c} / N$) is the number
of polymers per unit volume, and
\begin{equation}
\label{eq:ctilde}
\frac{\tilde{c}}{N} = \frac{N_\text{p}}{V}
\end{equation}
where, $N_\text{p}$ is the total number of polymers in the system, and $V$ is the
system volume. Since the correlation blobs are space filling, this implies,
\begin{equation*}
\xi_\text{c}^{3} N_\text{c} N_\text{p} = V
\end{equation*}
or, from Eq.~(\ref{eq:ctilde}),
\begin{equation}
\label{eq:ctildebyN}
\frac{\tilde{c}}{N} = \frac{1}{\xi_\text{c}^{3} N_\text{c}}
\end{equation}
Solving for $\xi_\text{c}^{3}$ we get,
\begin{equation}
\label{eq:xicube}
\xi_\text{c}^{3} = \left( \frac{N}{N_\text{c}} \right) \frac{1}{\tilde{c}}
\end{equation}
At the overlap concentration, since $R_\text{eq} = b N^{\nu}$, where $b$ is the size of a monomer, and $\nu$ is the Flory exponent, we have,
\begin{equation}
\label{eq:ctildestar}
\tilde{c}^{*} \sim \frac{N}{R_\text{eq}^{3}} \sim b^{-3} N^{1 - 3\nu}
\end{equation}
Multiplying and dividing the right hand side of Eq.~(\ref{eq:xicube})
with $\tilde{c}^{*}$, and using Eq.~(\ref{eq:ctildestar}), we get
\begin{equation}
\label{eq:xicube2}
\xi_\text{c}^{3} = \left( \frac{N}{N_\text{c}} \right) \left( \frac{\tilde{c}^{*}}{\tilde{c}} \right) b^{3} N^{3\nu - 1}
\end{equation}
For semidilute solutions, it can be shown that (see supplementary material
in \citet{Jain2012a}),
\begin{equation}
\label{eq:nc}
N_\text{c} = \left( \frac{\tilde{c}}{\tilde{c}^{*}} \right)^{\tfrac{1}{3\nu - 1}}
\end{equation}
As a result,
\begin{equation*}
\label{eq:xic3}
\xi_\text{c}^{3} = b^{3} N^{3\nu} \left( \frac{\tilde{c}}{\tilde{c}^{*}} \right)^{-1} \left( \frac{\tilde{c}}{\tilde{c}^{*}} \right)^{\tfrac{-1}{3\nu - 1}}
\end{equation*}
which simplifies to,
\begin{equation}
\label{eq:xicube3}
\xi_\text{c}^{3} = b^{3} N^{3\nu} \left( \frac{c}{c^{*}} \right)^{\tfrac{- 3\nu}{3\nu - 1}}
\end{equation}
where, we have used (\ccs) in place of $(\tilde{c} / \tilde{c^{*}})$ since they are
identical ($c = \tilde{c} \mk$, where \mk\ is the mass of a monomer).

The diffusivity of a correlation blob is given by
\begin{equation}
\label{eq:diffcorrblob}
\mathcal{D}_\text{c} = \frac{\kB T}{\zeta_\text{c}}
\end{equation}
where, $\zeta_\text{c}$ is the friction coefficient for the blob. Since the chain
segment within a correlation blob obeys Zimm dynamics,
\begin{equation}
\label{eq:zetac}
\zeta_\text{c} \sim \etas \xi_\text{c}
\end{equation}
and as a result,
\begin{equation}
\label{eq:diffcorrblob2}
\mathcal{D}_\text{c} = \frac{\kB T}{\etas \xi_\text{c}}
\end{equation}
The relaxation time for a single blob can be derived from the expression,
\begin{equation*}
\tau_\text{c} = \frac{\xi_\text{c}^{2}}{\mathcal{D}_\text{c}} = \frac{\xi_\text{c}^{2}}{(\kB T / \etas \xi_\text{c})}
\end{equation*}
or,
\begin{equation}
\label{eq:chainrelax1}
\tau_\text{c} = \frac{\etas \xi_\text{c}^{3}}{\kB T}
\end{equation}
If we define,
\begin{equation}
\label{eq:monorelax}
\tau_{0} = \frac{\etas b^{3}}{\kB T},
\end{equation}
as the monomer relaxation time, it follows from Eq.~(\ref{eq:xicube3}) that,
\begin{equation}
\label{eq:chainrelax2}
\tau_\text{c} = \tau_{0} N^{3\nu} \left( \frac{c}{c^{*}} \right)^{\tfrac{- 3\nu}{3\nu - 1}}
\end{equation}
Since a single chain in a semidilute solution is equivalent to a Rouse chain
of correlation blobs in a melt, it follows that the relaxation time of the
chain as a whole, $\tau_\text{chain}$ is given by~\citep{RubCol03}
\begin{equation*}
\tau_\text{chain} = \tau_\text{c} N_\text{c}^{2}
\end{equation*}
or,
\begin{equation}
\label{eq:chainrelax3}
\tau_\text{chain} = \left( \frac{\etas \xi_\text{c}^{3}}{\kB T} \right) N_\text{c}^{2}
\end{equation}

\section{ \label{pincblob} The magnitude of the stretching force and the size of Pincus blobs in shear flow}

It is straightforward to determine, from Eq.~(\ref{eq:forceblob}), the stretching force on a DNA chain at which the size of a Pincus blob becomes of order of the Kuhn step length. At 20$\degC$, since the thermal energy $k_\text{B}T$ is roughly 4 pN nm, and $b_\text{K} = 100$ nm, the threshold force $f_\text{max}$ at which $\xi_{S} \ge b_\text{K}$ is of order 0.04 pN. When $f = f_\text{max}$, the chain is a fully stretched blob pole of Pincus blobs, with each blob consisting of a single Kuhn step. As a result, the notion of the screening of monomer-monomer interactions has no relevance for  stretching forces $f \ge f_\text{max}$.  It is therefore important to obtain an estimate of the typical stretching force on a DNA chain  in shear flow, along with an estimate of the magnitude of the  Pincus blob size relative to the size of a correlation blob, and the Kuhn step size. It is worth noting that in the context where Pincus blobs are usually  discussed, namely in the measurement of the force-extension behaviour of DNA, 0.04 pN is an extremely small force, well within the linear regime, while non-linear force-extension behaviour is typically observed for forces of $\mathcal{O}(1)$ pN and greater~\cite{Marko19958759}.

\begin{table*}[t]
\addtolength{\tabcolsep}{-2pt}
\caption{Estimate of the magnitude of various quantities associated with Pincus blobs that arise in the shear flow of semidilute $\lambda$-phage DNA solutions (correlation blob size $\xi_\text{c}$, number of correlation blobs in a chain $N_\text{c}$, the size $\xi_\text{S}$ and number $X$ of Pincus blobs, the number of correlation blobs in a Pincus blob $m$, and the stretching force on a chain $f$), as a function of scaled concentration $c/c^*$, and non-dimensional shear rate $\lambda_\eta \dot \gamma$. The temperature is maintained constant at 25 $\degC$ for the estimation of these values.}
\label{tab:pincblobshear}
\vspace{10pt}
\centering
\setlength{\tabcolsep}{4pt}
{\def\arraystretch{1.1}
\begin{tabular}{ l  l l l l c  c c c c }
\hline
$c/c^*$ & $\lambda_\eta$ (s) &  $\xi_\text{c}$ (nm)  &  $N_\text{c}$ & $\dot \gamma$ (s$^{-1}$) & $\lambda_\eta \dot \gamma$ & $\xi_\text{S}$ (nm) & $m$ &$f$ (pN) & $X$ \\
\hline
\hline
\multirow{6}{*}{1.74 } & \multirow{6}{*}{0.358} & \multirow{6}{*}{1153.1}  & \multirow{6}{*}{2}  & 0.01  & 3.58E-3  & 4496.4 & 15.20 &0.001 & 0.1\\
&  & 	 &  & 0.1 & 3.58E-2 & 2528.5 & 4.81 & 0.002 & 0.4  \\ 
&   &  &  & 1 & 0.358 & 1421.9 & 1.52 & 0.003 & 1  \\ 
&   &  &  & 10 & 3.58 & 799.6 & 0.48 & 0.005 & 4  \\
&   &  &  & 100  & 35.8    & 449.6  & 0.15  & 0.009 & 13  \\
&   &  &  & 1000 & 358     & 252.8  & 0.05  & 0.016 & 42  \\  
\hline \hline 
\multirow{6}{*}{2.72 } & \multirow{6}{*}{0.481} & \multirow{6}{*}{930.6}  & \multirow{6}{*}{3}  & 0.01 & 0.00481 & 4135.1 & 19.75 & 0.001 & 0.2\\
    &       &        &    & 0.1  & 0.0481  & 2325.3 & 6.24  & 0.002 & 1   \\
     &       &        &    & 1    & 0.481   & 1307.6 & 1.97  & 0.003 & 2   \\
     &       &        &    & 10   & 4.81    & 735.3  & 0.62  & 0.006 & 6   \\
     &       &        &    & 100  & 48.1    & 413.5  & 0.20  & 0.010 & 18  \\
     &       &        &    & 1000 & 481     & 232.5  & 0.06  & 0.018 & 56  \\ 
     \hline 
\hline 
\multirow{6}{*}{4.35 } & \multirow{6}{*}{0.776} & \multirow{6}{*}{742.8}  & \multirow{6}{*}{6}  & 0.01 & 0.00776 & 3786.6 & 25.99 & 0.001 & 0.2\\
     &       &        &    & 0.1  & 0.0776  & 2129.4 & 8.22  & 0.002 & 1   \\
     &       &        &    & 1    & 0.776   & 1197.4 & 2.60  & 0.003 & 2   \\
     &       &        &    & 10   & 7.76    & 673.4  & 0.82  & 0.006 & 8   \\
     &       &        &    & 100  & 77.6    & 378.7  & 0.26  & 0.011 & 24  \\
     &       &        &    & 1000 & 776     & 212.9  & 0.08  & 0.019 & 76  \\ 
     \hline 
\hline
\multirow{6}{*}{6.85} & \multirow{6}{*}{2.347} & \multirow{6}{*}{597.3}  & \multirow{6}{*}{11}  & 0.01 & 0.02347 & 3477.5 & 33.89 & 0.001 & 0.3 \\
     &       &        &    & 0.1  & 0.2347  & 1955.6 & 10.72 & 0.002 & 1   \\
     &       &        &    & 1    & 2.347   & 1099.7 & 3.39  & 0.004 & 3   \\
     &       &        &    & 10   & 23.47   & 618.4  & 1.07  & 0.007 & 10  \\
     &       &        &    & 100  & 234.7   & 347.8  & 0.34  & 0.012 & 33  \\
     &       &        &    & 1000 & 2347    & 195.6  & 0.11  & 0.021 & 103 \\ 
\hline
\hline
\end{tabular}
}
\end{table*}

In order to obtain estimates of the stretching force and Pincus blob size in shear flow, it is necessary to calculate the value of the equilibrium end-to-end distance $R_\text{0,eq}$ under dilute conditions. \citet{Pan2014b} have estimated that in a dilute solution,  at 25$\degC$, the length of a chain segment of $\lambda$-phage DNA within a thermal blob is roughly of order $10^4$ base pairs. Since the length of a single base pair is 0.34 nm, it follows that the length of the chain segment $\ell_\text{T}$ within a thermal blob is 3400 nm. With random walk statistics obeyed within a thermal blob, the size of the thermal blob is $\xi_\text{T} = \sqrt{\ell_\text{T} b_\text{k}} = 583$ nm, the number of monomers within the thermal blob is $g = (\xi_\text{T}/ b_\text{k})^2 \approx 34$, and the total number of thermal blobs in the chain is $N_\text{T} = N_\text{k}/g = 165/34 \approx 5$. The thermal blobs exclude each other, and the chain of thermal blobs obeys self-avoiding walk statistics. Consequently, using the value $\nu = 0.6$ for the Flory exponent, and the values of $N_\text{T}$ and $\xi_\text{T}$ derived above, implies that at 25$\degC$, the dilute solution equilibrium end-to-end vector is roughly, $R_\text{0,eq} = N_\text{T}^\nu \, \xi_\text{T} \approx 1530$ nm. 

It is worth estimating the stretching force at which the Pincus blob size is equal to the dilute solution equilibrium end-to-end vector, since that would provide a lower bound on the stretching force below which the entire chain would be within a Pincus blob. From Eq.~(\ref{eq:forceblob}), and using a value of 4 pN nm for $k_\text{B}T$, the minimum force $f_\text{min}$ at which $\xi_{S} \approx R_\text{0,eq}$ is roughly 0.003 pN. It follows that for stretching forces that lie in the range $f_\text{min} =0.003 \,\, \text{nm} < f <  f_\text{max} = 0.04 \,\, \text{nm}$, the Pincus blob size is between $R_\text{0,eq}$ and $b_\text{k}$. The question that remains to be addressed is, in shear flow, what range of shear rates correspond to this range of stretching forces. 

For any given value of $c/c^*$, the number of correlation blobs in a chain $N_\text{c}$ (see Eq.~(\ref{eq:nc})), and the size of the correlation blob $\xi_\text{c}$ (see Eq.~(\ref{eq:xicube3})), are determined once and for all. In terms of $R_\text{0,eq}$ and $N_\text{c}$, the latter equation for $\xi_\text{c}$ can be rewritten  as,
\begin{equation}
\label{eq:xicappc}
\xi_\text{c} = R_\text{0,eq} \, N_\text{c}^{- \nu}
\end{equation}
Solving Eq.~(\ref{eq:sizeblob1}) for  $m$, the number of correlation blobs in a Pincus blob is,
\begin{equation}
\label{eq:numcorblob}
m = \left( \frac{\xi_\text{S}}{\xi_\text{c}} \right)^{2}
\end{equation}
Substituting for $m$ from Eq.~(\ref{eq:numcorblob}) and for $\xi_\text{c}$ from Eq.~(\ref{eq:xicappc}) into Eq.~(\ref{eq:m2}), gives the following equation for the size of a Pincus blob as a function of $c/c^*$ and shear rate $\dot \gamma$,
\begin{equation}
\label{eq:xiSappc}
\xi_\text{S} =  \left[ \frac{\kB T}{\etas \dot{\gamma}} \frac{R_\text{0,eq}}{N_\text{c}^{\nu}}\right]^{\tfrac{1}{4}}
\end{equation}
Once the size of a Pincus blob is determined from Eq.~(\ref{eq:xiSappc}), the stretching force $f$ can be determined from Eq.~(\ref{eq:forceblob}), and the number of Pincus blobs $X$ in a chain can be obtained from  Eq.~(\ref{eq:X}). As a result, all quantities associated with Pincus blobs that are formed due to the shear flow of semidilute $\lambda$-phage DNA solutions can be determined from experimentally measurable quantities. DNA solutions prepared in a Tris-Ethylenediaminetetraacetic (TE) acid buffer have a viscosity that is indistinguishable from water. Noting that, in units relevant to the present discussion, water has a viscosity of $10^{-9}$ pN nm$^{-2}$ s, the various blob scaling quantities discussed above have been calculated and tabulated in Table~\ref{tab:pincblobshear}. 

It is clear from Table~\ref{tab:pincblobshear} that for the range of values of $c/c^*$ considered, there are roughly two orders of magnitude in shear rate, roughly between $\dot \gamma = 0.1$ and $\dot \gamma = 10$, in which  $f_\text{min} \lesssim f \lesssim  f_\text{max}$, with the Pincus blob size $\xi_\text{S}$ lying between $\xi_\text{c}$ and $R_\text{0,eq}$, and approximately between $\mathcal{O}(1)$ to $\mathcal{O}(10)$ Pincus blobs formed on a chain. Since relaxation times are of $\mathcal{O}(1)$ second, the relevant Weissenberg numbers are also in this range. It is also apparent that even at significantly higher shear rates, $\xi_\text{S} \ge b_\text{k}$, and $f < f_\text{max} = 0.04$ pN. It appears that the drag force on a chain due to the surrounding fluid undergoing simple shear flow, does not give rise to stretching forces of the same orders of magnitude as observed in force-extension experiments. This is reflected in the weak dependence of the stretching force on shear rate, $f \sim \dot \gamma^{1/4}$, which can be derived by combining Eq.~(\ref{eq:xiSappc}) and Eq.~(\ref{eq:forceblob}). 

As mentioned earlier, for shear rates at which $\xi_\text{S}$ is smaller than $\xi_\text{c}$, the scaling arguments derived in section~\ref{subsec:colbyrelax} would have to be revised since several of the underlying assumptions would no longer be valid. A careful examination of the interplay between different blob length scales that are present in a semidilute solution subjected to extensional flow has been carried out recently in Ref.~\cite{Prabhakar16}. For the purposes of the present paper, however, we note that firstly the magnitudes of stretching forces and blob sizes derived within scaling theory are approximate estimates, since all numerical pre-factors are ignored, and secondly, the main goal of the scaling analysis in section~\ref{subsec:colbyrelax} is the derivation of the appropriate dependence of the large scale relaxation time on the scaled concentration $c/c^*$ (see  Eq.~(\ref{eq:relaxcolby})), which has been shown in section~\ref{subsec:rouse} to be extremely useful in obtaining universal data collapse.

\end{document}